\documentclass[prb,a4paper,floatfix,nofootinbib]{revtex4-1}
\usepackage{amsmath}
\usepackage{fancybox}
\usepackage{eepic}
\usepackage{times}
\usepackage{latexsym}
\usepackage{pifont}
\usepackage{graphicx}
\usepackage{epstopdf}
\usepackage{bm}

\def\rr{\boldsymbol{r}}
\def\GG{\boldsymbol{G}}
\def\qq{\boldsymbol{q}}
\def\kk{\boldsymbol{k}}

\begin{document}
\title{Accurate optical properties from first principles: a Quasiparticle Self consistent $GW$ plus Bethe-Salpeter Equation approach} 
\author{Brian Cunningham}
\address{School of Mathematics and Physics, Queen's University Belfast, Belfast BT7 1NN, Northern Ireland, United Kingdom}
\author{Pooya Azarhoosh}
\address{King's College London, 
        London WC2R 2LS, United Kingdom} 
\author{Dimitar Pashov}
\address{King's College London, 
        London WC2R 2LS, United Kingdom} 
\author{Myrta Gr\"uning}
\address{School of Mathematics and Physics, Queen's University Belfast, Belfast BT7 1NN, Northern Ireland, United Kingdom} 
\author{Mark van Schilfgaarde}
\address{King's College London, 
        London WC2R 2LS, United Kingdom} 
\begin{abstract}
  We present an approach to calculate the optical absorption spectra that combines the quasiparticle self-consistent $GW$ method [Phys. Rev. B, {\bf 76} 165106 (2007)] for the electronic structure with the solution of the ladder approximation to the Bethe-Salpeter equation for the macroscopic dielectric function. The solution of the Bethe-Salpeter equation has been implemented within an all-electron framework, using a linear muffin-tin orbital basis set, with the contribution from the non-local self-energy to the transition dipole moments (in the optical limit) evaluated explicitly. 
  This approach addresses those systems whose electronic structure is poorly described within the standard perturbative $GW$ approaches with as a starting point density-functional theory calculations. The merits of this approach have been exemplified by calculating optical absorption spectra of a strongly correlated transition metal oxide, NiO, and a narrow gap semiconductor, Ge. In both cases, the calculated spectrum is in good agreement with the experiment.
  It is also shown that for systems whose electronic structure is well-described within the standard perturbative $GW$, such as Si, LiF and {\it h}-BN, the performance of the present approach is in general comparable to the standard $GW$ plus  Bethe-Salpeter equation. It is argued that both vertex corrections to the electronic screening and the electron-phonon interaction are responsible for the observed systematic overestimation of the fundamental bandgap and spectrum onset.
\end{abstract}
\pacs{42.25.Bs,11.10.St,71.15.-m,78.20.-e}
\maketitle

%
%
%
\section{Introduction} 
The interplay between computer simulation and experiment may prove key for envisaging `new' materials that can be used as components in the technological devices of the future.\cite{graphene1,graphene2,graphene3}   Characterization of the interactions between electrons in a solid and the interaction with external perturbations is rapidly progressing due to advances in theory, experimental techniques and computational power and resources.  Many different theoretical methods exist for calculating the electronic structure in a material.  One very successful and widely used theory is density functional theory (DFT).\cite{PhysRev.136.B864,kohnsham}  It is well understood, however, that DFT has many drawbacks, for example, being a theory which is exact for the ground state, its description of excited states is only approximate. The approximation for the exchange-correlation energy functional, such as the local density approximation (LDA), introduces further problems (see e.g. Ref.~\onlinecite{GW_aryasetiawan}).

Approaches that combine DFT with Many-Body perturbation theory have become widely used over the past decade to treat excited states and spectroscopic properties of materials. For charged excitations, the $GW$ approximation ($GW$A), based on the work of Hedin,\cite{hedin} has proved very successful in calculating the quasi-particle electronic structure in solids.\cite{GW_aryasetiawan} For optical properties, starting from the quasi-particle electronic structure, the  Bethe-Salpeter equation (BSE)\cite{BSE_paper,onida_electronic_2002} accurately introduces the two-particle electron-hole interactions---through the ladder diagrams---that are essential to describe the excitonic effects which dominates e.g. the optical absorption of semiconductors and insulators.

The most commonly used form of the $GW$A is $G_0W_0$ (also referred to as one-shot or single-shot $GW$).\cite{louie}  In this approach the single particle Green's function and polarization are constructed from the DFT (usually within the LDA or Generalized Gradient Approximation) energies and eigenfunctions.
The polarization determines the screened Coulomb interaction $W$ and the self-energy is then calculated from the Green's function and $W$; hence the name $GW$.  The electronic structure from the DFT calculation is then {\em perturbatively} corrected by replacing the contribution to the energy eigenvalues from the DFT exchange-correlation potential with the contribution from the diagonal part of the self-energy.\cite{louie}  Though the $GW$A has been very successful in calculating the band gaps of semi-conductors and insulators,\cite{GW_aryasetiawan}
it also has several drawbacks. In particular there is marked dependence on the DFT starting point,\cite{SC_BE} and it has been long known that LDA-based $GW$ systematically underestimates bandgaps in simple semiconductors.\cite{QSGW_PRL}  Difficulties are particularly severe in narrow gap semiconductors, such as CuInSe$_2$,\cite{Vidal10} for which the DFT gap is often inverted. As a consequence of the poor description of the electronic structure, optical properties are also poorly described.  

The $GW$A is an approximation to a formally exact formalism developed by Hedin,\cite{hedin} where a set of five coupled equations are to be solved self-consistently. Though the full self-consistent solution of Hedin's equation cannot be achieved, some form of self-consistency may seem as the natural way to improve over the  $G_0W_0$ approach. Different forms of self-consistency have been introduced. The most straighforward self-consistency is to replace the corrected eigenvalues\cite{SC_ener} in either $G$ and/or $W$\footnote{see for example, Ref.~\onlinecite{SC_BE}, where results are presented for $G_0W_0$, $GW_0$ and $GW$} or self-consistency in the energies and not the eigenfunctions. More sophisticated forms of self-consistency---such as the form employed in this work---involve as well the eigenfunctions. In general, in spite of the additional computational effort---which is substantial in case of the self-consistency on the eigenfunctions--- different forms of self-consistency may not improve systematically on the $G_0W_0$ approach. The homogeneous electron gas\cite{SC_EG} and spectral functions in transistion metals\cite{Belashchenko06} are well-known examples where self-consistency gives a worse result than the $G_0W_0$.  At least for Jellium, the next higher order diagram approximately restores the 1-shot $GW$ result.\cite{Shirley96}  Reference~\onlinecite{QSGW_paper} (Appendix A) has traced the one main reason for this difficulty to the imperfect cancellation of the renormalization factor $Z$. In many cases, such as CuInSe$_2$\cite{Vidal10} and the transition metal oxides,\cite{QSGW_paper} self-consistency on the eigenfunctions is critical to get the correct electronic structure and, as a consequence, to calculate the optical properties of materials, as it has been shown by Bruneval {\it et al} in Ref~\onlinecite{Bruneval06b} for the dielectric response of Cu$_2$O. 

In this work, we present a first principles framework and computational tool to calculate the dielectric function (Sec.~\ref{ss:diefun}), and hence the optical properties, of materials for which the $G_0W_0$ approach (Sec.~\ref{ss:g0w0}) provides a poor description of the electronic structure. By following a strategy similar to  Ref~\onlinecite{Bruneval06b}, in the proposed framework, the electronic structure is calculated with the quasiparticle self-consistent $GW$ (QS$GW$) method.\cite{QSGW_prl1,QSGW_PRL,QSGW_paper}  In the QS$GW$, the `best' starting Hamiltonian (as opposed to the usual DFT one) is determined using the $GW$A iteratively. The new starting point is chosen so that the quasiparticles (i.e., the single particle eigenfunctions and eigenenergies) generated from the effective one-particle DFT-like potential match the quasiparticles generated from the $GW$. (Sec.\ref{ss:qsgw})

The electronic structure obtained with this method---which was already implemented in the code Questaal~\cite{questaal_web}---is then used to calculate the dielectric function from the solution of the BSE---which has been newly implemented in the same code  (Sec.~\ref{ss:bse}). This approach is referred in the following as QS$GW$+BSE.
We detail how the BSE has been numerically implemented within an all-electron framework using a linear muffin-tin orbital basis set (Sec.~\ref{ss:lmto}).
We also discuss the calculation of the non-local contribution to the transition dipole moments, which are a key ingredient to obtain the dielectric function. Usually, the transition dipole moments are calculated using the DFT electronic structure,\cite{opt_mat_el,opt_mat_el2} or by rescaling the QSGW moments by the ratio of DFT and $GW$ eigenenergy differences. This approach will be adequate when the DFT eigenfunctions give a good description of the electronic structure, however, it cannot be used when the DFT bandgaps are inverted or too small. We then employed here an approach to obtain the non-local contribution to the transition dipole moments explicitly (Sec.~\ref{subsec:ome}).

The QS$GW$+BSE approach is then assessed by calculating the optical absoprtion of prototypical systems (Sec.~\ref{sec:res}). First, we test and assess the approach for Si, LiF and bulk hexagonal BN, that are systems where the widely used plane wave pseudopotential $G_0W_0$ method\cite{espresso,yambo,vasp} works relatively well. We then turn to Ge and NiO, two systems for which we show is critcal to introduce self-consistency into the $GW$.
%
%
\section{Theory and approximations}
\subsection{Dielectric function}\label{ss:diefun}
To obtain optical properties, the key quantity is the frequency-dependent macroscopic dielectric function $\epsilon_{\rm M}(\omega)$ which is defined as the optical (long wavelength) limit ($\boldsymbol{q}\rightarrow 0$) of the inverse of the macroscopic average ($\boldsymbol{G}=\boldsymbol{G}'=0$) of the inverse dielectric matrix, $\epsilon^{-1}$, in Fourier space:\footnote{In all the following we are considering infinite crystals so that any function $f(\rr)$ having crystalline symmetries can be represented in Fourier space as $f(\qq+\GG)$, where $\GG$s are reciprocal lattice vectors.} 
\begin{equation}\label{eq:macro}
\epsilon_{\rm M}(\omega)=\lim_{\boldsymbol{q}\rightarrow 0}\frac{1}{\epsilon_{\boldsymbol{G}=\boldsymbol{G}'=0}^{-1}(\boldsymbol{q},\omega)}.
\end{equation}

The inverse dielectric matrix is defined as the functional derivative of the total potential with respect to the external potential, $\epsilon^{-1}(1,2)=\delta V_{\rm tot}(1)/\delta V_{\rm ext}(2)$ (with $1=(\boldsymbol{r}_1,t_1,\sigma_1)$) and can be expressed as
\begin{equation}
\epsilon^{-1}(1,2)=\delta(1,2)+\int \text{d}3\, v(1,3)\chi(3,2).
\label{eq:epsm1}
\end{equation}
In Eq.~\eqref{eq:epsm1}, we introduced the reducible polarizability, $\chi(1,2)=\delta\rho_{\rm ind}(1)/\delta V_{\rm ext}(2)$,  which describes the change induced in the electronic density due to the external potential. 
Similarly, the dielectric matrix is given by, 
\begin{equation}
\epsilon(1,2)=\delta(1,2)-\int \text{d}3\, v(1,3)P(3,2),
\end{equation}
where $P(1,2)=\delta\rho_{\rm ind}(1)/\delta V_{\rm tot}(2)$ is the irreducible polarizability, which describes the change induced in the electronic density due to the total potential. 

 It can be shown\cite{Hanke_mod} that the macroscopic dielectric function can be calculated from a modified response function, $\bar{P}$, through the equation
\begin{equation}
\epsilon_{\rm M}(\omega)=1-\lim_{\boldsymbol{q}\rightarrow 0}v_{\boldsymbol{G}=0}(\boldsymbol{q})\bar{P}_{\boldsymbol{G}=\boldsymbol{G}'=0}(\boldsymbol{q},\omega),
\end{equation}
where $v_{\boldsymbol{G}}(\boldsymbol{q})=4\pi/|\boldsymbol{q}+\boldsymbol{G}|^2$ is the Coulomb interaction in Fourier space.
The modified response function for optical absorption is related to the irreducible polarizability through the equation
\begin{equation}\label{eq:modified_resp}
\bar{P}=P+P\bar{v}\bar{P},~~~~ \bar{v}_{\boldsymbol{G}}(\boldsymbol{q})=\displaystyle\left\{\displaystyle\begin{array}{cl}0 & {\rm if~}\boldsymbol{G}=0 \\ \vspace{-0.2cm}\\ \displaystyle\frac{4\pi}{|\boldsymbol{q}+\boldsymbol{G}|^2}&{\rm otherwise.}\end{array}\right. 
\end{equation}

 The simplest expression for $P$ is the random phase approximation (RPA),\cite{onida_electronic_2002} which assumes a sum over independent particle transitions\footnote{The RPA polarization presented is essentially Fermi's golden rule} and in frequency space is given by:
   \small
 \begin{equation}\label{eq:RPA_pol}
P_{\rm RPA}(\boldsymbol{r},\boldsymbol{r}';\omega)=\sum_{n_1n_2}(f_{n_2}-f_{n_1})\frac{\psi_{n_2}^{*}(\boldsymbol{r})\psi_{n_1}(\boldsymbol{r})\psi_{n_1}^{*}(\boldsymbol{r}')\psi_{n_2}(\boldsymbol{r}')}{\varepsilon_{n_2}-\varepsilon_{n_1}-\omega-{\rm i}\eta},
\end{equation}\normalsize
where $\varepsilon_{n}$, $\psi_n$ and $f_n$  are the single-particle energies, wavefunctions and occupations (note that the state index, $n_i$, contains the band, $k$-point and spin indices) and $\eta$ a small positive number.\footnote{$\eta$ effectively makes the imaginary part of $P_{\rm RPA}$ a Dirac-delta function that ensures energy conservation.}  The choice of $\varepsilon_{n}$ is discussed in the next subsections.



\subsection{Electronic Structure: DFT+$GW$}\label{ss:g0w0}
The electronic structure, $\varepsilon_{n}$, $\psi_n$ and $f_n$, is needed as an input to calculate the irreducible polarizability and thus the macroscopic dielectric function.
The computationally cheapest way to obtain $\varepsilon_{n}$, $\psi_n$ and $f_n$ from first-principles is within the Kohn-Sham DFT framework, which corresponds to the self-consistent solution of a set of Schr\"odinger-like equations with the single-particle Hamiltonian
\begin{equation}
H_0(\boldsymbol{r})=-\frac{1}{2}\boldsymbol{\nabla}^2+V_{\rm ext}(\boldsymbol{r})[\rho]+V_H(\boldsymbol{r})[\rho]+V_{\rm XC}(\boldsymbol{r})[\rho].
\label{eq:oneHam}
\end{equation}
Besides  $V_{\rm ext}(\boldsymbol{r})$, the external potential due to the nuclei and any external fields, the  Hartree potential, $V_H(\boldsymbol{r})$, and the exchange-correlation potential, $V_{\rm XC}(\boldsymbol{r})$, appear in Eq.~\eqref{eq:oneHam}. The former describes the classical mean-field electron-electron interaction; the latter potential contains the missing correlation effects in some given approximation (see e.g. Ref.~\onlinecite{MARQUES20122272}).
Though the  Kohn-Sham DFT band structure $\varepsilon_{n}$ is usually in qualitative agreement with the quasiparticle band structure, the band gaps obtained  from the $\varepsilon_{n}$ are known to be underestimated by about 40\% due to both the neglection of the derivative discontinuity and the approximation for $V_{\rm XC}(\boldsymbol{r})$.~\cite{PhysRevB.37.10159, doi:10.1063/1.2189226}  
To obviate this problem, the state-of-the-art is to combine DFT with Green's function theory in what is usually referred to as the DFT+$GW$ approach (see e.g. Refs.~\onlinecite{onida_electronic_2002,AULBUR20001}). In the latter, the  $\varepsilon_{n}$ obtained from the solution of the Kohn-Sham DFT equations are perturbatively corrected at the first order:
\begin{equation}
E_{n\boldsymbol{k}}=\varepsilon_{n\boldsymbol{k}}+\langle\psi_{n\boldsymbol{k}}|\Sigma^{GW}(E_{n\boldsymbol{k}})-V_{\rm XC}|\psi_{n\boldsymbol{k}}\rangle.
\label{eq:gwperb}
\end{equation}
In Eq.~\eqref{eq:gwperb}, $\Sigma^{GW}$ is the self-energy in the so-called $GW$ approximation.\cite{hedin,GW_aryasetiawan} The general expression for the self-energy, and related quantities, is given by:
\begin{align}
  \Sigma(1,2)&={\rm i}\int{\rm d}(34)~G(1,3^{+})W(1,4)\Lambda(3,2,4)\label{eq:selfE1}\\
  G(1,2)&=G_{0}(1,2) + \int{\rm d}(34)~G_{0}(1,3)\Sigma(3,4)G(4,2)\label{eq:GreenF}\\
 W(1,2)&=\int{\rm d}3~\epsilon^{-1}(1,3)v(3-2)\label{eq:scrpot}\\
  \Lambda(1,2,3)&=\delta(1,2)\delta(1,3) + \nonumber \\&~~~\int {\rm d}(4567)\frac{\delta \Sigma(1,2)}{\delta G(4,5)} G(4,6) G(7,5)\Lambda(6,7,3)\label{eq:vert}
\end{align}
where $G$ is the Green's function,  $W$ is the screened Coulomb interaction---with $\epsilon^{-1}$ the inverse dielectric function introduced in Eq.~\eqref{eq:epsm1}, and $\Lambda$ is the irreducible vertex function. This set of equations [\eqref{eq:selfE1}--\eqref{eq:vert}], known as Hedin's equations,~\cite{PhysRev.139.A796,schwinger,schwinger2} is completed by the equation for the irreducible polarizability (needed to determine $\epsilon^{-1}$):
\begin{equation}
P(1,2)=-{\rm i}\int {\rm d}(34)~G(1,3)\Lambda(3,4,2)G(4,1^{+}).\label{eq:irrpol}
\end{equation}
In Eqs.~\eqref{eq:selfE1} and~\eqref{eq:irrpol}, the $+$ superscript implies $t'=t+\eta$.

The $GW$ approximation to the self-energy corresponds to approximate {\em(a)} the vertex as $\Lambda(1,2,3) \approx \delta(1,2)\delta(1,3)$, and {\em(b)} the Green's function by the noninteracting Green's function (in frequency space and subsuming the spin and band indices into a single index $n_i$)
\begin{equation}\label{eq:green}
G_{0}(\boldsymbol{r},\boldsymbol{r}',\omega)=\sum_{n\boldsymbol{k}}\frac{\psi_{n\boldsymbol{k}}(\boldsymbol{r})\psi_{n\boldsymbol{k}}^{*}(\boldsymbol{r}')}{\omega-\varepsilon_{n\boldsymbol{k}}\pm{\rm i}\eta}.
\end{equation}
As a consequence of {\em(a)} and {\em(b)} the inverse microscopic dielectric matrix in the expression for $W$ [Eq.~\eqref{eq:scrpot}]  is calculated within the RPA [Eq.~\eqref{eq:RPA_pol}].

Equation~\eqref{eq:gwperb} is nonlinear as the self-energy on the RHS depends on $E_{n\boldsymbol{k}}$. Usually Eq.~\eqref{eq:gwperb} is linearized as:
\begin{equation}
E_{n\boldsymbol{k}}=\varepsilon_{n\boldsymbol{k}}+Z_{n\boldsymbol{k}}\langle\psi_{n\boldsymbol{k}}|\Sigma(\varepsilon_{n\boldsymbol{k}})-V_{\rm XC}|\psi_{n\boldsymbol{k}}\rangle
\label{eq:gwlin}
\end{equation}
where the renormalization factor $Z_{n\kk}$ reads:
\begin{equation}
Z_{n\boldsymbol{k}}=(1-\displaystyle \partial\Sigma(\omega)/\partial\omega|_{\omega=\varepsilon_{n\boldsymbol{k}}})^{-1}.
\label{eq:Zren}
\end{equation}
Though in standard $GW$ calculations the renormalization factor $Z$ is usually employed in Eq.\eqref{eq:gwlin}, there are several arguments for setting the $Z$-factor equal to $1$.  One argument relies on the $Z$-factor cancellation in the expression for the self-energy (for details, see Appendix A of Ref.~\cite{QSGW_paper}). Another argument relies on the formula for the derivative discontinuity of the DFT-RPA functional\cite{deriv_disc}, which is the same expression in Eq.~(\ref{eq:gwlin}), but for the $Z$ factor being equal to $1$. In this work we adopt the $Z=1$ choice and we show that indeed this generally leads to a better agreement with experimental results.
\subsection{Electronic Structure: QS$GW$}\label{ss:qsgw}
The above DFT+$GW$ approach gives a perturbative correction to the Kohn-Sham DFT energies at the first order. At this order, the wavefunctions are not corrected. As a consequence, the DFT+$GW$ approach works well when the Kohn-Sham DFT gives already a reasonable, physically correct description of the electronic structure and properties of the system.
When this is not the case, some form of self-consistency is usually introduced into the method. The simplest form of self-consistency is to replace the corrected energy $E_n$ [Eq.~\eqref{eq:gwlin}] either in the Green's function [Eq.~\eqref{eq:green}], or in the RPA polarization [Eq.~\eqref{eq:RPA_pol}] entering the screened potential $W$, or in both. Again, in this scheme the wavefunctions are not corrected, so this form of self-consistency is not expected to work well when DFT gives a wrong physical description of the system (e.g. predicts a metal rather than an insulator). In those cases, one needs more sophisticated approaches which provide improved wavefunctions. Existing approaches include starting from hybrid DFT---as e.g. in  Ref.~\onlinecite{PSSB:PSSB200945204}---or the Coulomb-hole screened exchange approximation for the self-energy,~\cite{PhysRev.139.A796}---as e.g. in Ref.~\onlinecite{PhysRevLett.99.266402}---and using the QS$GW$ approach,\cite{QSGW_paper} which is the method of choice of this work.

In the QS$GW$ approach once the self-energy has been calculated within the $GW$ approximation, rather than correcting the Kohn-Sham energies as in Eq.~\eqref{eq:gwlin}, a new effective single-particle potential is determined with,\cite{QSGW_paper}
\begin{equation}
\label{eq:QSGW_Vxc}
\begin{array}{rl}
\bar V_{\rm XC}=\frac{1}{2}\sum_{n_1n_2}|\psi_{n_1}\rangle\left\{\right.&{\rm Re}[\Sigma^{GW}(\varepsilon_{n_1})]+\\
&\left.{\rm Re}[\Sigma^{GW}(\varepsilon_{n_2})]\right\}_{n_1n_2}\langle\psi_{n_2}|,\end{array}
\end{equation} 
where $\Sigma^{GW}_{n_1n_2} = \langle \psi_{n_1}|\Sigma^{GW}|\psi_{n_2}\rangle$. This expression for $\bar V_{\rm XC}$ effectively minimizes the perturbation in Eq.~\eqref{eq:gwperb}.~\footnote{This is not the only possible choice. For example an alternative expression for $\bar V_{\rm XC}$ exists as detailed in Ref~\onlinecite{QSGW_paper}} 

Then, by substituting $V_{\rm XC}$ with $\bar V_{\rm XC}$ in Eq.~\eqref{eq:oneHam}, a new set of single-particle energies and wavefunctions can be determined. In turn, those can be used to re-calculate the $GW$ self-energy, and the whole procedure can be repeated  until self-consistency in the energies and eigenvalues is achieved. 
The main advantage of this procedure is that the resulting electronic structure does not depend on the quality of the Kohn-Sham DFT electronic structure for the system.

\subsection{The Bethe-Salpeter equation}\label{ss:bse}
 


An approximation for the irreducible polarizability, which improves over the RPA, can be obtained if in the expression for the vertex, Eq.~(\ref{eq:vert}), we assume that $\delta \Sigma/\delta G={\rm i}W$ (i.e., we ignore the vertex in Eq.~(\ref{eq:selfE1}) when calculating $\delta\Sigma/\delta G$).\footnote{The vertex in Eq.~(\ref{eq:selfE1}) can been shown to effectively cancel with the $Z$-factor, see for example Appendix~A in Ref.~\cite{QSGW_paper}} Then we can arrive at an expression for the polarization, $-{\rm i}GG\Lambda$.
This results in $P\approx P^0-P^0WP$, where $P^0$ is the RPA polarization.
When inserting this expression in the definition for the modified response function in Eq.~(\ref{eq:modified_resp}), we obtain $\bar P\approx  P^0+P^0K\bar P$ with the kernel $K = \bar{v}-W$.\footnote{For parmagnetic systems with one spin channel treated explicitly, if the occupancies are 1 then the kernel becomes $2\bar{v}-W$ (see Ref.~\onlinecite{onida_electronic_2002}).  This is because only singlet excitations contribute to optical absorption.  Examining the expressions for the kernel: for $W$ we must have $\sigma=\sigma'$ where $\sigma(\sigma')$ is the spin of states $n_1$ and $n_2$ ($n_3$ and $n_4$), however this is not the case for $\bar{v}$.}
In a 4-point polarization representation: 
\begin{align}
\bar{P}(1234)&=P^{0}(1234) \notag\\ &+\int{\rm d}(5678)P^{0}(1256)K(5678)\bar{P}(7834), \\
  K(1234)&=\delta(12)\delta(34)\bar{v} -\delta(13)\delta(24)W(12) \label{eq:bsepol}
\end{align}
and $P^0(1212)=P^0(12)$. As an additional approximation, the kernel is usually assumed to be static. In few works this approximation has been relaxed, see e.g. Ref.~\onlinecite{PhysRevLett.91.176402}.

The Dyson-like equation for the polarizability is usually transformed in a eigenproblem for an effective 2-particle Hamiltonian by introducing the basis of single particle eigenfunctions which diagonalize the RPA polarization. 
Using the completeness of the eigenfunctions, any 4-point quantity can be expanded as 
\small\begin{align}
S(\boldsymbol{r}_1,\boldsymbol{r}_2,\boldsymbol{r}_3,\boldsymbol{r}_4)=\hspace{-0.15cm}\sum_{\tiny n_1n_2 n_3n_4\normalsize}& S_{n_1n_2n_3n_4}\times\nonumber\\ &\psi_{n_1}(\boldsymbol{r}_1)\psi_{n_2}^{*}(\boldsymbol{r}_2)\psi_{n_3}^{*}(\boldsymbol{r}_3)\psi_{n_4}(\boldsymbol{r}_4),
\end{align}\normalsize
where we have again combined band, spin and wavevector indices, and $S_{n_1n_2n_3n_4}=\int{\rm d}(\boldsymbol{r}_1\boldsymbol{r}_2\boldsymbol{r}_3\boldsymbol{r}_4) S(\boldsymbol{r}_1,\boldsymbol{r}_2,\boldsymbol{r}_3,\boldsymbol{r}_4)\times$ $\psi_{n_1}^{*}(\boldsymbol{r}_1)\psi_{n_2}(\boldsymbol{r}_2)\psi_{n_3}(\boldsymbol{r}_3)\psi_{n_4}^{*}(\boldsymbol{r}_4)$.\vspace{0.1cm}\par
 
Inserting the expression for the RPA polarization from Eq.~\eqref{eq:RPA_pol} in Eq.~\eqref{eq:bsepol}, one arrives at the following expression for the polarization 
\begin{equation}
P_{\scriptsize\begin{array}{l}n_1n_2\boldsymbol{k}\\n_3n_4\boldsymbol{k}'\end{array}}(\boldsymbol{q},\omega)=\left[H(\boldsymbol{q})-\omega\right]_{\scriptsize\begin{array}{l}n_1n_2\boldsymbol{k}\\n_3n_4\boldsymbol{k}'\end{array}}^{-1}(f_{n_4\boldsymbol{k}'+\boldsymbol{q}}-f_{n_3\boldsymbol{k}'}),
\end{equation}
whereby the conservation of momentum we have $\boldsymbol{k}_{2(4)}=\boldsymbol{k}_{1(3)}+\boldsymbol{q}$; and
\begin{equation}\begin{array}{rl}
H_{\scriptsize\begin{array}{l}n_1n_2\boldsymbol{k}\\n_3n_4\boldsymbol{k}'\end{array}}(\boldsymbol{q})=&(\varepsilon_{n_2\boldsymbol{k}'+\boldsymbol{q}}-\varepsilon_{n_1\boldsymbol{k}'})\delta_{n_1n_3}\delta_{n_2n_4}\delta_{\boldsymbol{k}\boldsymbol{k}'}-\\&(f_{n_2\boldsymbol{k}+\boldsymbol{q}}-f_{n_1\boldsymbol{k}})K_{\scriptsize\begin{array}{l}n_1n_2\boldsymbol{k}\\n_3n_4\boldsymbol{k}'\end{array}}(\boldsymbol{q}).\end{array}
\end{equation}
The expression $(H-\omega)^{-1}$ in the spectral representation is:
\begin{equation}
\left[H(\boldsymbol{q})-\omega\right]_{ss'}^{-1}=\sum_{\lambda\lambda'}\frac{A_{s}^{\lambda}(\boldsymbol{q})N_{\lambda,\lambda'}^{-1}(\boldsymbol{q})A_{s'}^{*\lambda'}(\boldsymbol{q})}{E_\lambda(\boldsymbol{q})-\omega\pm{\rm i}\eta},\label{eq:Hminone}
\end{equation}
where $A_{s}^{\lambda}(\boldsymbol{q})$ is element $s=n_1n_2\boldsymbol{k}$ of the eigenvector of $H(\boldsymbol{q})$ with corresponding eigenvalue $E_\lambda(\boldsymbol{q})$ and $N$ is the overlap matrix. When the Tamm-Dancoff approximation is adopted,~\cite{TD_myrta} $H$ is Hermitian and Eq.~\eqref{eq:Hminone} reduces to $\displaystyle\sum_{\lambda}\displaystyle\frac{A_{s}^{\lambda}(\boldsymbol{q})A_{s'}^{*\lambda}(\boldsymbol{q})}{E_\lambda(\boldsymbol{q})-\omega\pm{\rm i}\eta}$.\par
Finally, the macroscopic dielectric function is calculated as
\begin{align}\label{eq:macro_exp}
\epsilon_{\rm M}(\omega)=1-&\lim_{\boldsymbol{q}\rightarrow 0}\frac{8\pi}{|\boldsymbol{q}|^2\Omega N_kN_{\sigma}}\times \notag \\ &\sum_{ss'}\Delta f_{s'}(\boldsymbol{q})\rho_{s}(\boldsymbol{q})\left[H(\boldsymbol{q})-\omega\right]_{ss'}^{-1}\rho_{s'}^{*}(\boldsymbol{q}),
\end{align}
where $\Omega$, $N_k$ and $N_{\sigma}$ are the cell volume, number of $k$-points in the full Brillouin zone and number of spin channels treated explicitly; $\Delta f_{s'}(\boldsymbol{q})=(f_{n_4\boldsymbol{k}'+\boldsymbol{q}}-f_{n_3\boldsymbol{k}'})$ and
\begin{equation}
\rho_{s}(\boldsymbol{q})=\langle\psi_{n_2\boldsymbol{k}+\boldsymbol{q}}|e^{{\rm i}\boldsymbol{q}\cdot\boldsymbol{r}}|\psi_{n_1\boldsymbol{k}}\rangle \label{eq:rhoq}
\end{equation}
are the transition dipole matrix elements, often also referred as oscillators.

\section{Numerical implementation}

\subsection{Evaluation of the kernel matrix elements}\label{ss:lmto}
Our numerical implementation of the BSE relies on a linear muffin-tin orbital basis.\cite{methfessel_lmto,QSGW_paper,fusion}  The eigenfunctions are expanded in Bloch-summed muffin-tin orbitals in spheres around atom centers.  The radial part of the eigenfunctions in these spheres is expanded by numerical solutions of the radial Schr\"{o}dinger equation.  In the region between the spheres, the eigenfunctions are then expanded in either smoothed Hankel functions\cite{fusion} and/or plane waves.  Expanding the interstitial in plane waves, the eigenfunctions are
\begin{equation}\label{eq:eig_exp}
\Psi_{n\boldsymbol{k}}(\boldsymbol{r})=\sum_{\boldsymbol{R}u}\alpha_{\boldsymbol{R}u}^{\boldsymbol{k}n}\varphi_{\boldsymbol{R}u}^{\boldsymbol{k}}(\boldsymbol{r})+\sum_{\boldsymbol{G}}\beta_{\boldsymbol{G}}^{\boldsymbol{k}n}P_{\boldsymbol{G}}^{\boldsymbol{k}}(\boldsymbol{r}),
\end{equation}
where $\boldsymbol{R}$ denotes the atomic site and $u$ is a composite index that contains the angular momentum of the site along with an index that denotes either: a numerical solution of the radial Schr\"{o}dinger equation at some representative energy; its energy derivate (since the energy dependence has been linearized by expanding in a Taylor series about the representative energy\cite{OKA}); or a local orbital which is a solution at an energy well above or below the representative energy.  In $GW$ and the BSE a basis is required that expands the product of eigenfunctions.  Expanding the interstitial in plane waves, the product eigenfunctions will also be expanded in plane waves, and within the spheres the basis is expanded by $\varphi_{Ru}(\boldsymbol{r})\times\varphi_{Ru'}(\boldsymbol{r})$.  This mixed product basis (MPB) is denoted $M_I^{\boldsymbol{k}}(\boldsymbol{r})$.

In the MPB, the two components of the kernel $K$ in Eq.~\eqref{eq:bsepol} read as 
\small\begin{equation}\label{kernel_eq}\begin{array}{lrl}
\displaystyle \bar{v}_{{\scriptsize\begin{array}{l}n_1n_2\boldsymbol{k}\\n_3n_4\boldsymbol{k}'\end{array}\normalsize}}(\boldsymbol{q})&=\displaystyle\sum_{I,J}&\langle\psi_{n_2,\boldsymbol{k}+\boldsymbol{q}}|\psi_{n_1,\boldsymbol{k}}\widetilde{M}_I^{\boldsymbol{q}}\rangle \bar{v}_{IJ}(\boldsymbol{q})\\&&\times\langle\widetilde{M}_J^{\boldsymbol{q}}\psi_{n_3,\boldsymbol{k}'}|\psi_{n_4,\boldsymbol{k}'+\boldsymbol{q}}\rangle\\&&\\
\displaystyle W_{{\scriptsize\begin{array}{l}n_1n_2\boldsymbol{k}\\n_3n_4\boldsymbol{k}'\end{array}\normalsize}}(\boldsymbol{q})&=\displaystyle\sum_{I,J}&\langle\psi_{n_3,\boldsymbol{k}'}|\psi_{n_1,\boldsymbol{k}}\widetilde{M}_I^{\boldsymbol{k}'-\boldsymbol{k}}\rangle W_{IJ}(\boldsymbol{k}'-\boldsymbol{k};\omega=0)\\&&\times\langle\widetilde{M}_J^{\boldsymbol{k}'-\boldsymbol{k}}\psi_{n_2,\boldsymbol{k}+\boldsymbol{q}}|\psi_{n_4,\boldsymbol{k}'+\boldsymbol{q}}\rangle.
\end{array}
\end{equation}\normalsize
The macroscopic part of $v$ is set to zero, Eq.~(\ref{eq:modified_resp}), by diagonalizing $v_{IJ}$, i.e., transforming to the basis in Ref.~\onlinecite{kotani_MPB} and setting $v_{\mu}=0$ before transforming back to $v_{IJ}$.

\subsection{Optical matrix elements}\label{subsec:ome}
To calculate the macroscopic dielectric function in Eq.~\eqref{eq:macro_exp}, the transition dipole matrix elements (TDME) [Eq.~\eqref{eq:rhoq}]
in the limit $\boldsymbol{q}\rightarrow 0$ are needed.  Those can be evaluated numerically, e.g. by employing the  offset-$\Gamma$ method, whereby these matrix elements are determined for finite $\boldsymbol{q}$ near zero. Alternatively, one can expand Eq.~\eqref{eq:rhoq} in a Taylor series about $\qq = 0$ and truncate to the first order, leaving ${\rm i}\boldsymbol{q}\cdot\langle\psi_{n_2\boldsymbol{k}}|\boldsymbol{r}|\psi_{n_1\boldsymbol{k}}\rangle$.

Since the position operator $\hat{\boldsymbol{r}}$ is ill-defined when periodic boundary conditions are imposed, 
  the commutation relation $[\hat{H},\hat{\boldsymbol{r}}]={\rm i}\boldsymbol{\nabla}$, which holds when only local potentials appear in the Hamiltonian, is used to obtain the relation for the TDME
\begin{equation}\label{eq:opt_mat}
{\rm i}\boldsymbol{q}\cdot\langle\psi_{n_2\boldsymbol{k}}|\boldsymbol{r}|\psi_{n_1\boldsymbol{k}}\rangle_{\rm Loc}=-\boldsymbol{q}\cdot\frac{\langle\psi_{n_2\boldsymbol{k}}|\boldsymbol{\nabla}|\psi_{n_1\boldsymbol{k}}\rangle}{\varepsilon_{n_2\boldsymbol{k}}-\varepsilon_{n_1\boldsymbol{k}}}.
\end{equation}
However, the effective Hamiltonian corresponding to Green's function methods contains the non-local self energy operator. The usual way to account for the contribution from the non-local self-energy is to replace $(\varepsilon_{n_2\boldsymbol{k}}-\varepsilon_{n_1\boldsymbol{k}})$ with $(\varepsilon_{n_2\boldsymbol{k}}^{\rm LDA}-\varepsilon_{n_1\boldsymbol{k}}^{\rm LDA})$,\cite{opt_mat_el2,opt_mat_el} which corresponds to rescaling the local contribution by a factor $\displaystyle\frac{(\varepsilon_{n_2\boldsymbol{k}}-\varepsilon_{n_1\boldsymbol{k}})}{(\varepsilon_{n_2\boldsymbol{k}}^{\rm LDA}-\varepsilon_{n_1\boldsymbol{k}}^{\rm LDA})}$. 

This approach is exact when a simple scissor operator is applied to correct the LDA eigenvalues. Otherwise, it is an approximation that works well when the LDA eigenfunctions approximate well the quasiparticle ones and it is expected to fail in the case of e.g., NiO, where the LDA is inaccurate; or for Ge where the LDA predicts a semi-metal and thus for some $\boldsymbol{k}$ the energy difference between the bottom conduction (BC) and the top valence (TV) band, $(\varepsilon_{{\rm BC},\boldsymbol{k}}^{\rm LDA}-\varepsilon_{{\rm TV},\boldsymbol{k}}^{\rm LDA})$, can be zero or negative.\par
In this work we account for the contribution from the non-local self energy by explicitly calculating matrix elements of the velocity operator \cite{opt_mat_el}
\begin{equation}\label{eq:dsigdp}
\boldsymbol{v}=\boldsymbol{p}-{\rm i}\frac{\partial\Sigma(\boldsymbol{r},\boldsymbol{p})}{\partial \boldsymbol{p}},
\end{equation}
where $\Sigma(\boldsymbol{r},\boldsymbol{p})=\int{\rm d}\boldsymbol{r}'\Sigma(\boldsymbol{r},\boldsymbol{r}')e^{{\rm i}(\boldsymbol{r}-\boldsymbol{r}')\cdot{\boldsymbol{p}}}$; which can be derived from the commutation between the Hamiltonian and position, and using the translation operator $\hat{T}(\boldsymbol{x})\psi(\boldsymbol{r})=\psi(\boldsymbol{r}+\boldsymbol{x})$.

\newcommand{\bfk}{{\bf k}}
\newcommand{\bfr}{{\bf r}}
\def\Psikn{\Psi_{n{\bf k}}}
\def\vxc{V^{\rm xc}}
\def\brl{{\bf R}L}
\def\brlp{{{\bf R}'L'}}
\def\iDelta{{\it \Delta}}

In QS$GW$, $\Sigma$ is replaced by its static approximation, Eq.~\ref{eq:QSGW_Vxc}.
$\bar V_{\rm XC}({\bf k})$ is calculated in the eigenfunction basis, 
and the LDA potential subtracted.  In this way the difference can be conveniently 
added to the LDA hamiltonian.  Call this difference
$\iDelta\vxc_{nm}({\bf k}) = \left(\bar V_{\rm XC}-V^{\rm LDA}_{\rm XC}\right)_{nm}$.

The eigenfunctions (see Eq.~\ref{eq:eig_exp}) can be expressed in the general form
\begin{eqnarray}
\label{eq:lmtopsi}
\Psikn(\bfr) = \sum_{\brl{j}} z^{{\bf k}n}_{\boldsymbol{R}u} \chi^{\bfk}_{\boldsymbol{R}u}({\bf r})
\end{eqnarray}
where, for a particular band $n$, $\Psikn(\bfr)$ is defined by the (eigenvector) coefficients $z^{{\bf k}n}_{\boldsymbol{R}u}$ and the shape of the $\chi^{\bfk}_{\boldsymbol{R}u}({\bf r})$. The basis functions are augmented smoothed Hankel functions defined by smoothing radius and energy, or a local orbital (see Section IIA in Ref.~\onlinecite{QSGW_paper} for details). When performing the perturbative $GW$ approximation, the eigenfunctions are then expressed according to Eq.~\ref{eq:eig_exp} with the interstitial expanded in plane waves.

$\iDelta\vxc_{nm}({\bf k})$ can be rotated from the LMTO basis by
\begin{eqnarray}
\iDelta\vxc_{nm}({\bf k})=
\sum_{\boldsymbol{R}u,\boldsymbol{R}'u'}z^{{\bf{k}n}\dag}_{\boldsymbol{R}u}\,\iDelta\vxc_{\boldsymbol{R}u,\boldsymbol{R}'u'}({\bf k})\,z^{{\bf k}m}_{\boldsymbol{R}'u'}. 
 \nonumber \\
\label{eq:vxcfromvxcmto}
\end{eqnarray}
Because the method uses a real space basis,
$\iDelta\vxc_{nm}({\bf k})$ can be written as a Bloch sum
\begin{eqnarray}
\iDelta\vxc_{nm}({\bf k}) = 
\sum_{\bf T}e^{i{\bf k}\cdot{\bf T}}\iDelta\vxc_{\boldsymbol{R}u,{{\bf R'+T}u'}} .
\end{eqnarray}

The method computes $\iDelta\vxc_{\boldsymbol{R}u,{{\bf R'+T}}u'}$ 
on a regular mesh of points ${\bf k}_{\rm mesh}$, and inverting the process
\[
\iDelta \vxc_{nm}({\bf k}_{\rm mesh}) \to 
\iDelta\vxc_{\boldsymbol{R}u,\boldsymbol{R}'u'}({\bf k}_{\rm mesh})\to
\iDelta\vxc_{\boldsymbol{R}u,{{\bf R'+T}}u'}.
\]
as explained in Section IIG of Ref.~\onlinecite{QSGW_paper}.

Finally, the $\mathbf{k}$ derivative needed ($\mathbf{p}{=}\hbar\mathbf{k}$)
for Eq.~\ref{eq:dsigdp}, is readily computed by
differentiating the Bloch-summed form of $\iDelta \vxc_{nm}({\bf k}_{\rm mesh})$
with respect to $\mathbf{k}$.

%
%
\section{Results and Discussion}\label{sec:res}
We first assess the performance of QS$GW$+BSE for two prototypical systems: LiF (Sec.~\ref{ss:lif}) and Si (Sec.~\ref{ss:Si}). Then, we calculate the optical absorption spectrum of bulk {\it h}-BN (Sec.~\ref{ss:hbn}) for which $GW$+BSE calculations in the literature underestimate the position of the exciton peak and it has been suggested that some form of selfconsistency in the $GW$ calculations is needed.~\cite{PhysRevB.87.035404}
Finally, we calculate the optical absorption spectrum of Ge (Sec.~\ref{ss:Ge}) and NiO (Sec.~\ref{ss:nio}). For the former, DFT within the standard LDA/GGA predicts a direct semimetal rather than an indirect bandgap semiconductor; for the latter the DFT bandgap is ten times smaller than the experimental bandgap. Both systems justify the approach described in this work and highlight its strengths.

\subsection{Computational details}\label{ss:comp}
Table~\ref{tb:comp_details} contains the relevant  parameters used in the calculations. With the exception of {\it hexagonal}-BN ({\it h}-BN), the Bravais lattice of all systems considered are face-centered cubic.  In the $GW$ (single-shot and QS$GW$) the RPA polarization matrix is calculated by including all valence and a large number of the conduction states (between 50 and 100).  When calculating the spectrum within the RPA, the tetrahedron method\cite{QSGW_paper} is employed for integration over the Brillouin zone. For spectra calculated within the BSE, the broadening was applied according to Eq.~\ref{eq:Hminone} and varied to match experiment; except for in NiO where Gaussian broadening was applied to better agree with experiment. For LiF the broadening varies linearly. In the table we then report the broadening parameter at the onset and at the end of the considered energy range. More precisely, $\eta_\text{LiF}(\omega) = 0.053\omega - 0.57$, where $\omega$ is the photon energy in eV.

\begin{table}[h]
\centering
\label{tb:comp_details}
\begin{tabular}{r|c|c|c|c|c}\hline
&LiF & Si & {\it h}BN & Ge & NiO\\\hline\hline
$a$(\AA)&4.03&5.43&2.5 &5.66&4.17\\
$c$(\AA)&--&--&6.64&--&--\\
$G_{\rm Max}$(eV)&127&68&120&65&122\\
$N_{\bm k}$&12&16&10,10,5&12&8\\
$N_{v}$&4&4&6&4&11\\
$N_{c}$&4&4&8&5&6\\
$\eta$(eV)&0.07--0.7&0.14&0.2&0.2&0.27\\\hline
\end{tabular}
\caption{Parameters used in the calculations: lattice constant $a$ (and interlayer distance $c$ for {\it hexagonal}-BN); energy cut-off for the plane wave basis set $G_{\rm MAX}$; number of $\bm k$-points $N_{\bm k}$; the number of valence $N_v$ and conduction $N_c$ states used in the BSE; and the broadening $\eta$ used. Lorentzian broadening was used in all cases, except for in NiO, where Gaussian broadening was applied. Where two values are given, they refer respectively to the broadening at the spectrum onset and at the end of the considered energy range.}
\label{tb:comp_details}
\end{table}

When calculating the dielectric function within the BSE, due to the large memory and computational time requirements, we treat only a subset of transitions between valence and conduction bands at this level of theory (see Table~\ref{tb:comp_details}). Transitions to higher energy conduction bands (between 50 and 100) are included at the level of the RPA. 
 
\begin{figure}[h!]
\includegraphics[width=0.38\textwidth,clip=true,trim=0.0cm 0.0cm 0.0cm 0.0cm]{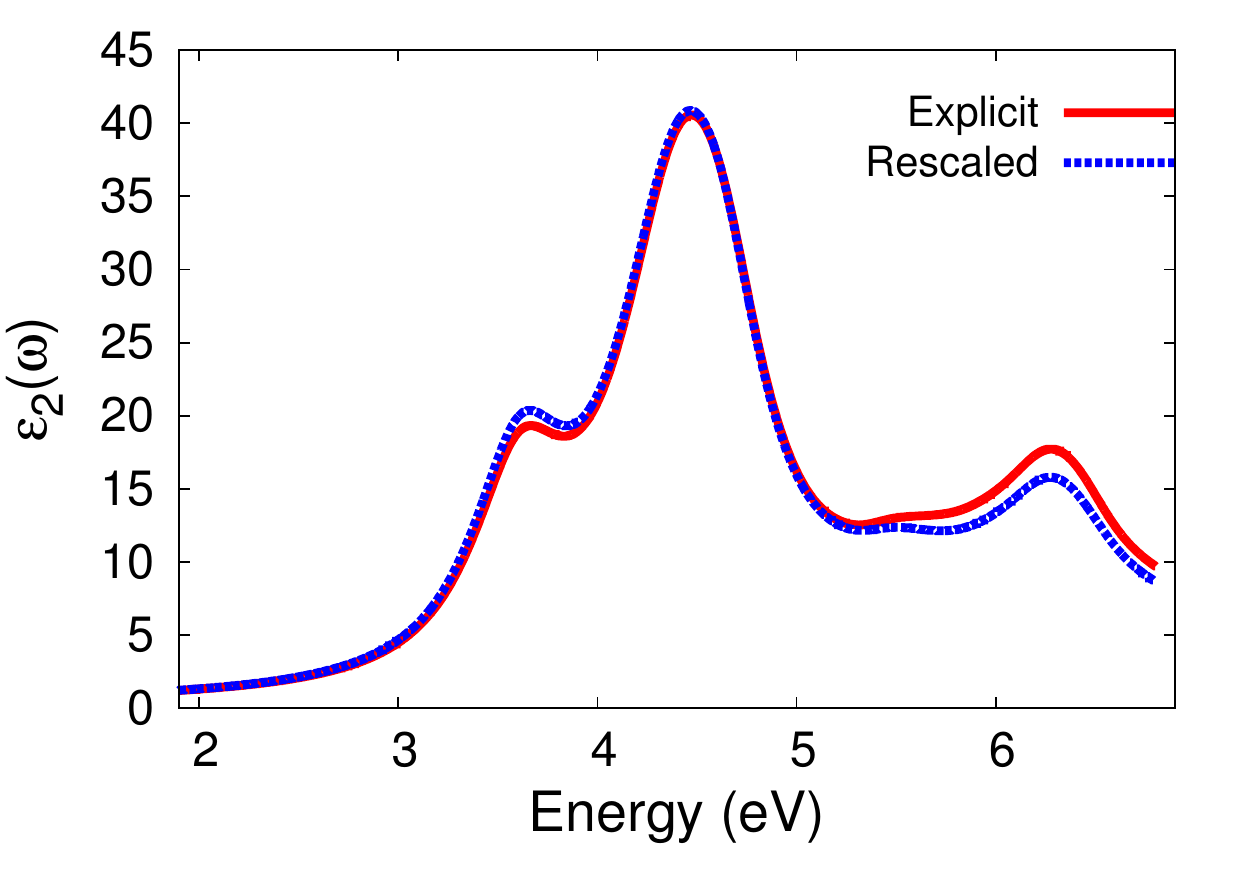}
\caption{Imaginary part of the macroscopic dielectric function for bulk Si. The spectra were calculated at the same level of theory (QS$GW$+RPA). Red continuous line: the nonlocal contribution to the TDMEs is calculated explicitly according to Eq.~(\ref{eq:dsigdp}); blue dashed line: the nonlocal contribution is accounted by rescaling by the ratio of the quasiparticle and DFT band gaps (see Sec.~\ref{subsec:ome} for details).}
\label{image:compare_si}
\end{figure}

\begin{figure}[h!]
\includegraphics[width=0.38\textwidth,clip=true,trim=0.0cm 0.0cm 0.0cm 0.0cm]{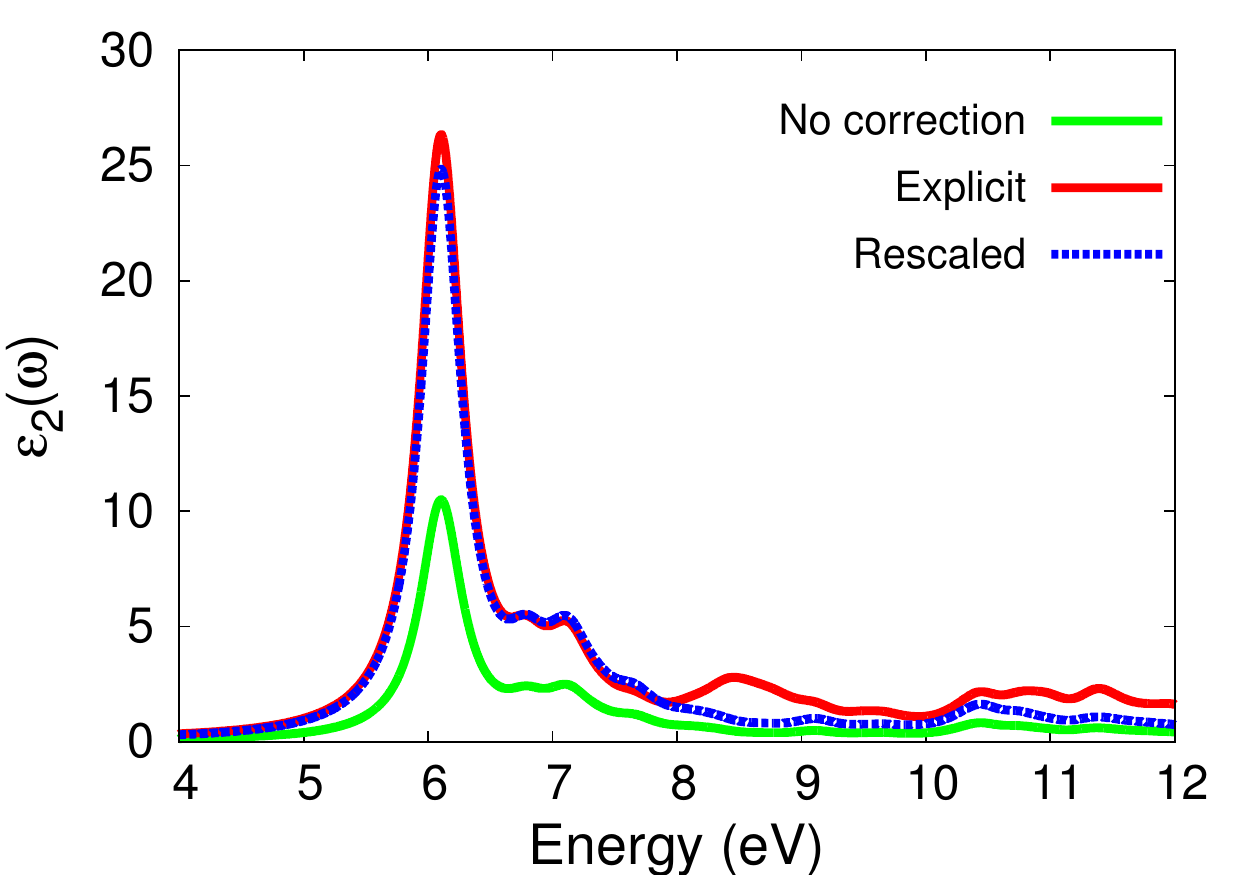}
\caption{Imaginary part of the macroscopic dielectric function for bulk {\it h}-BN. The spectra were calculated at the same level of theory (QS$GW$+BSE). Green continuous line: the nonlocal contribution to the TDMEs is neglected; red continuous line: the nonlocal contribution to the TDMEs is calculated explicitly according to Eq.~(\ref{eq:dsigdp}); blue dashed line: the nonlocal contribution is accounted by rescaling by the ratio of the quasiparticle and DFT band gaps. See Sec.~\ref{subsec:ome} for details.}
\label{image:compare_hbn}
\end{figure}

The nonlocal contribution to the TDMEs has been evaluated both according to Eq.~\eqref{eq:dsigdp} and by the bandgap rescaling discussed in Sec.~\ref{subsec:ome}. In Figs.~\ref{image:compare_si}--\ref{image:compare_hbn}, for silicon (at the QSGW+RPA level) and {\it h}-BN (at the QSGW+BSE level) we compare the evaluation of the TDMEs with the bandgap rescaling. In both cases, only marginal differences are observed. This is to be expected since for both silicon and {\it h}-BN the perturbative $G_0W_0$ approach is known to work well, meaning that the LDA wavefunctions are a good approximation to quasiparticle wavefunctions and the effect of quasiparticle corrections is approximately that of a scissor operator for which the  bandgap rescaling of the TDMEs is exact. 
For that reason, for LiF we use only the bandgap rescaling of TDMEs. For Ge and NiO, the bandgap rescaling cannot be used because of the inverted gap and the failure of the perturbative approach respectively. In that case,  TDMEs were evaluated only according to Eq.~\eqref{eq:dsigdp}. Finally, in Fig.~\ref{image:compare_hbn} we reported as well the spectrum obtained when the TDME is calculated without accounting for the nonlocal contribution from the self-energy. The intensity of the main features is reduced by about 50\% due to the sum rule violation. 

\subsection{Lithium Fluoride}
\label{ss:lif}
Lithium fluoride is a wide bandgap insulator with a strongly-bound charge-transfer exciton of the Frenkel type.~\cite{ISI:000258905700014} Here, it is considered as a prototypical system to test the validity of the approximations discussed and to assess the BSE implementation.
From thermoreflectance measurements~\cite{PhysRevB.13.5530} the fundamental bandgap of LiF at $\Gamma$ is estimated to be $14.2\pm 0.02$~eV.\footnote{Previous measurements on LiF reflection spectrum~\cite{Roessler:67} estimated the bandgap at $\Gamma$ to be $13.60\pm 0.06$ eV.} As to be expected, calculation of the electronic structure within DFT at the LDA level severely underestimates the fundamental bandgap (9.4 eV). Adding quasiparticle corrections within the $G_0W_0$ approximation gives a bandgap of 13.5 eV, when including the $Z$ renormalization factor in Eq.~\eqref{eq:Zren}, and of 14.2 eV when setting $Z=1$. These values are in good agreement with previous calculations at this level of theory (see e.g. Ref.~\onlinecite{lif_exp}).  As previously discussed in the literature (see e.g. Ref.~\onlinecite{QSGW_paper}), the success of the $G_0W_0$ approximation in predicting the bandgap of $sp$ semiconductors and insulators relies on error cancellation. Notably, on the one hand the screening potential is evaluated at the RPA level, missing the vertex corrections, leading to overestimating the bandgap; on the other hand the LDA energy differences which enter the RPA polarization are underestimated leading to overscreening, thus to underestimating the bandgap. Furthermore, it has been recently demonstrated\cite{phonons_walt} that inclusion of the electron-phonon interaction reduces the bandgap. The result we obtain at the QS$GW$ level is consistent with this picture: the bandgap is found to be 16~eV, thus substantially overestimated with respect to the experimental gap. This overestimation results from calculating the screening at the RPA level, but with quasiparticle energy differences, and from neglecting the electron-phonon interaction. 

\begin{figure}[h!]
\hspace{0.0cm}
\includegraphics[width=0.4\textwidth,clip=true,trim=0.0cm 0cm 0.0cm 0.0cm]{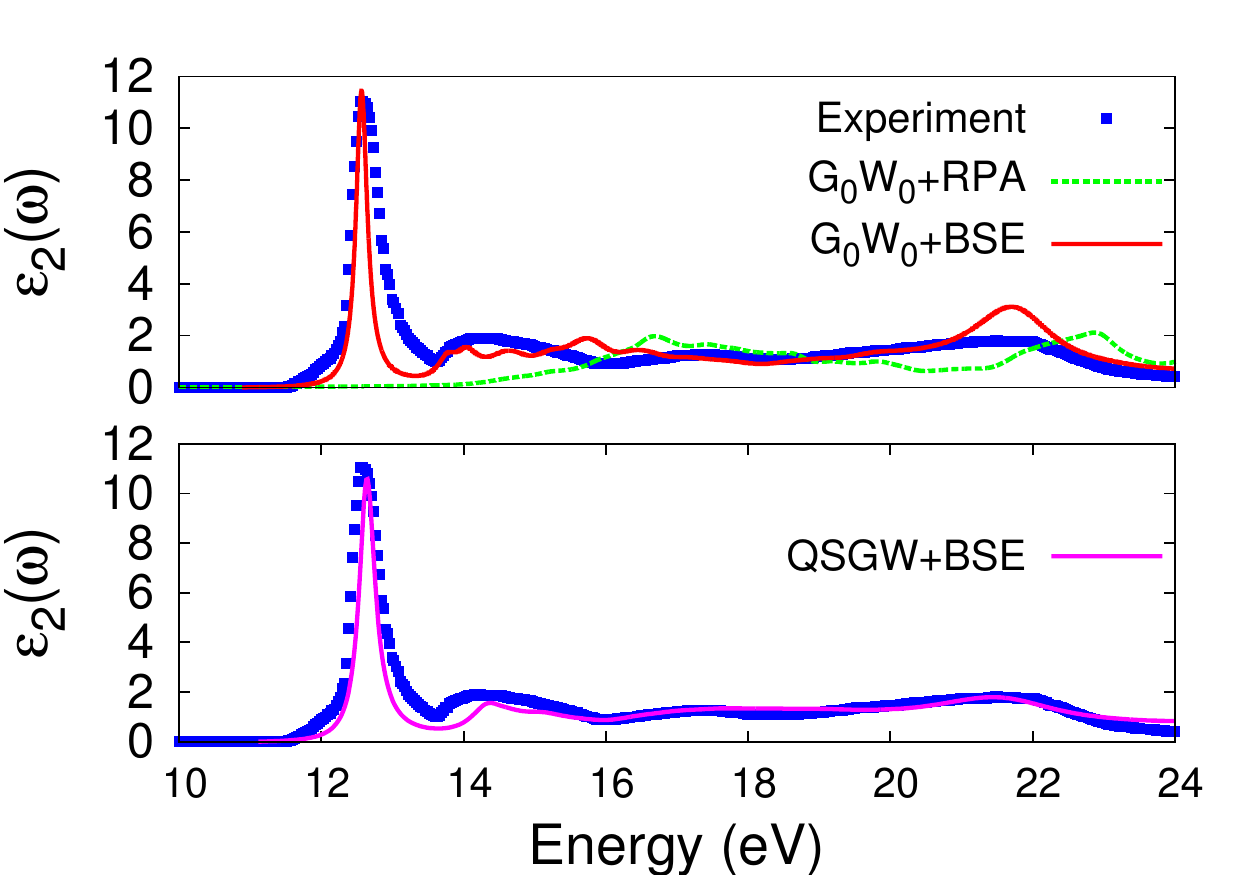}
\caption{Imaginary part of the macroscopic dielectric function for LiF. Upper panel: the experimental data (blue squares) \cite{Roessler:67} is compared with the results from the $G_0W_0$+RPA (green line) and the $G_0W_0$+BSE (red line). Lower panel: the experimental data (blue squares) is compared with the results from the QS$GW$+BSE. The spectrum is red-shifted by 0.9 eV to match the position of the first peak in the experimental spectrum.}
\label{image:lif}
\end{figure}
The results at $G_0W_0$ ($Z=1$) and QS$GW$ levels have then been used to obtain the macroscopic dielectric function within the BSE framework. Results for the imaginary part $\boldsymbol \varepsilon_2$ of the macroscopic dielectric function are compared in Fig.~\ref{image:lif} with the experimental absorption spectrum.~\cite{Roessler:67} The latter shows a sharp intense peak at about $12.6$~eV -- about $1.6$~eV below the fundamental band gap -- which has been identified as an exciton resonance. The position and intensity of the exciton resonance, and in general of all the absorption spectrum, is well reproduced at the $G_0W_0$+BSE level (top panel).  For a comparison when neglecting excitonic effects within $G_0W_0$+RPA, the theoretical spectrum onset is at about $14$ eV and the excitonic resonance is missing. From this result a binding energy of $1.7$~eV can be extracted; in very good agreement with the experimental results.~\cite{PhysRevB.13.5530}
These results are in agreement with the literature (see e.g. Refs~\onlinecite{lif_exp,yambo}) and validate (together with the results obtained for the other systems) the BSE implementation. The bottom panel shows the spectrum obtained at the QS$GW$+BSE level. The latter has been red-shifted by $0.9$~eV to match the position of the exciton in the experimental spectrum. The error in predicting the spectrum onset is due to the overestimation of the fundamental bandgap discussed above, which is only partially compensated by the overestimation of the exciton binding energy (about $2.4$~eV). The overestimations of bandgap and exciton binding energy originate both from underestimating the electronic screening in $W$. Apart from the spectrum onset, the overall shape of the spectrum is better reproduced within QS$GW$+BSE than within $G_0W_0$+BSE. In particular, the intensity of the spectral feature at about $22$~eV (assigned by Piancentini et al.~\cite{PhysRevB.13.5530} to an $X$ exciton) is well reproduced, while overestimated within the $G_0W_0$+BSE. It is worth to note that this is improvement is (at least) partly the effect of the larger broadening parameter used for the  QS$GW$+BSE spectrum. In fact, the broadening parameter was chosen to increase linearly with the photon energy (see Sec.~\ref{ss:comp}), so the QS$GW$+BSE spectrum, that is blue-shifted by almost 1 eV with respect to the $G_0W_0$+BSE, has a larger broadening parameter at the above-mentioned $X$ exciton feature ($\approx 0.65$~eV vs $\approx 0.60$~eV). 

\begin{table}[h]
\label{tb:seps}
\centering
\begin{tabular}{r|cc|cc|l}\hline
&\multicolumn{2}{|c|}{RPA}&\multicolumn{2}{|c|}{BSE}&\\ 
  &$G_0W_0$  & QS$GW$  & $G_0W_0$ & QS$GW$  &  Exp. \\\hline\hline
$\epsilon_{\infty}$ & 1.61 & 1.71 & 1.76 & 1.84 & 1.92 \\ \hline
\end{tabular}
\caption{Electronic part of the static dielectric constant, $\epsilon_{\infty}$, for LiF. Values at different levels of theory are compared with the experimental result.~\cite{lif_seps}}
\label{tb:seps}
\end{table}

Table~\ref{tb:seps} reports the values for the static dielectric constant, $\epsilon_{\infty}$ at the various levels of theory. Calculating the macroscopic dielectric function at the BSE level improved noticeably the agreement with the experimental value with respect to the RPA. When the electronic structure is calculated at the QS$GW$, rather than $G_0W_0$ level, the agreement with experiment is further improved ($1.84$ versus $1.92$).

\subsection{Silicon}
\label{ss:Si}
Silicon is a semiconductor which electronic structure and optical properties have been accurately characterized both theoretically and experimentally (see e.g. Refs.~\onlinecite{PhysRevB.47.2130, PhysRevB.36.4821}). For this reason it is often chosen as a prototypical system to assess approximations and test numerical implementation.
Table~\ref{tb:Sigap} summarizes the results for the fundamental bandgap -- which is indirect from the top of the valence in $\Gamma$ and the conduction band minimum (CBM), which occurs 85\% towards the boundary of the first Brillouin zone in the [100] direction -- and the minimum direct bandgap in $\Gamma$. The results follow the same trend observed for LiF. The underestimation of the LDA is partially corrected at the $G_0W_0$ level. A better agreement is obtained when the renormalization factor $Z$ in Eq.~\ref{eq:Zren} is set to $1$ taking into account cancellation in the expression for the self-energy.~\cite{QSGW_paper}
At the QS$GW$, the bandgap is slightly overestimated, as one would expect when neglecting vertex corrections and electron-phonon interactions. With respect to the wide-gap LiF, the QS$GW$ overestimation is relatively smaller, which can be expected as due to the larger screening, the vertex corrections are less important.    

\begin{figure}[h!]
\hspace{0.0cm}
\includegraphics[width=0.41\textwidth,clip=true,trim=0.0cm 0cm 0.0cm 0.0cm]{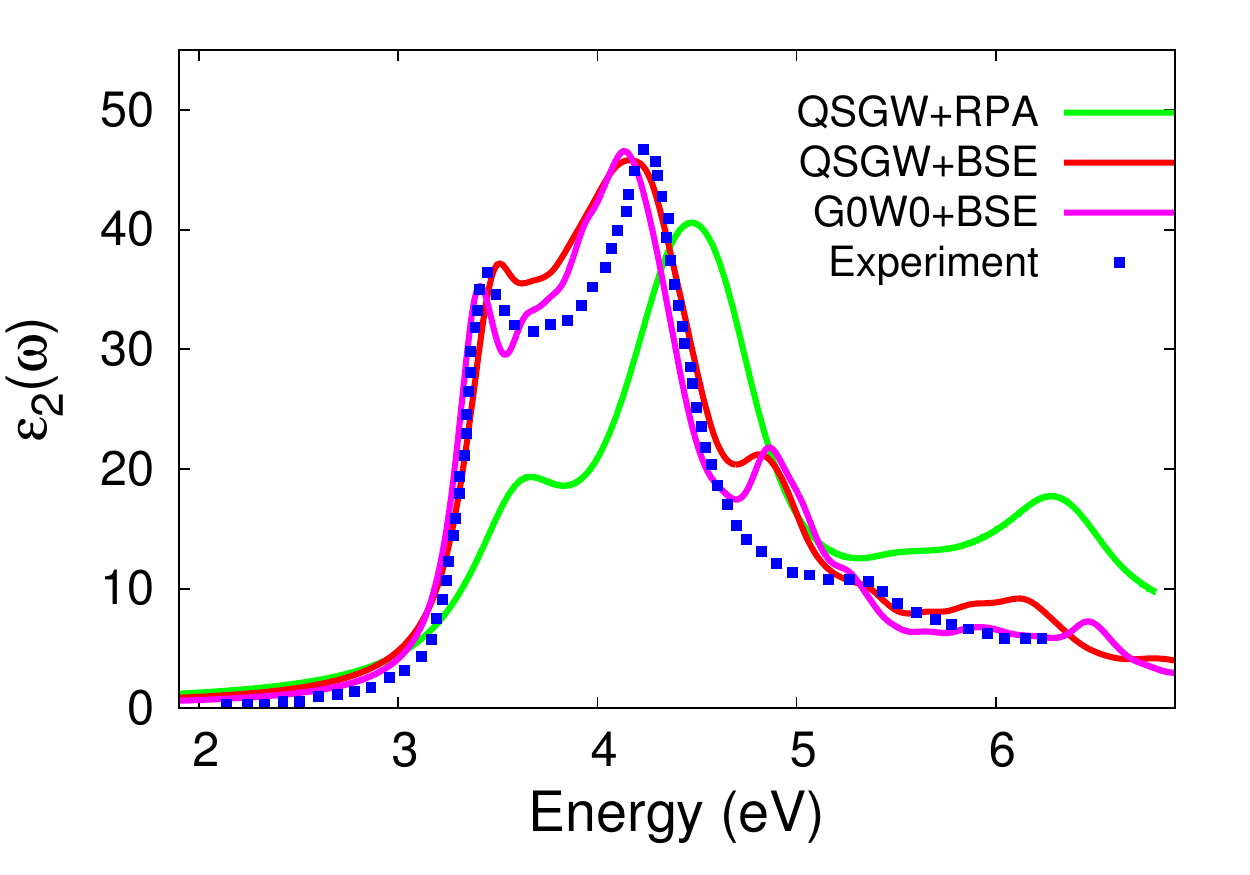}
\caption{Imaginary part of the macroscopic dielectric function for bulk Si. Theoretical results from $G_0W_0$+BSE (red line), QS$GW$+RPA (green line), and QSGW+BSE (purple line) are compared with the experimental data\cite{Si_exp2} (blue squares).}
\label{image:si}
\end{figure}

\begin{table}[h]
\centering
\label{tb:Sigap}
\begin{tabular}{r|cccc|l}\hline
  & LDA  & $G_0W_0$  & $G_0W_0\,(Z=1)$ & QS$GW$  &  Exp. \\\hline\hline
  $\Gamma-$CBM & 0.48 & 0.94 & 1.07 & 1.18 & 1.17  \\ \hline
  $\Gamma-\Gamma$ & 2.53 & 3.1  & 3.29  & 3.41 & 3.40 \\ \hline
\end{tabular}
\caption{Fundamental and minimum direct bandgap of Si at different levels of the theory and from experiment\cite{kittel} (all values in eV). See text.}
\label{tb:Sigap}
\end{table}
The imaginary part of the macroscopic dielectric function at the level of both $G_0W_0$+BSE and QS$GW$+BSE  is presented in Fig.~\ref{image:si}. Theoretical results are compared with the experimental spectrum~\cite{Si_exp2}.
As is well-known, the first peak is not well reproduced when excitonic effects are not taken into account as it can be seen from the QS$GW$+RPA results. Within the BSE level, the experimental spectrum is well reproduced, both using the electronic structure from $G_0W_0\,(Z=1)$ and QS$GW$, with minor differences.\par

\subsection{{\it hexagonal-}Boron Nitride}
\label{ss:hbn}
Bulk {\it h}-BN is a wide-gap layered semiconductor. The interest on this material is partly due to its similarity to graphite and to the possibility of obtaining few-layer compounds by exfoliation. As well, bulk {\it h}-BN has remarkable optical properties. For example, the strong excitonic features in the absorption spectrum~\cite{PhysRevLett.96.026402} or the high luminescence yield~\cite{Watanabe2004}.  Experimentally, the debate on a minimum direct or indirect bandgap has been solved only recently (see e.g. Refs~\onlinecite{Cassabois2016,Watanabe2004} and references therein) and the values for the fundamental bandgap obtained from different experiments cover a range of 3.5~eV.~\cite{SOLOZHENKO20011331,Watanabe2004} Furthermore, this discrepancy reflects as well in the interpretation of the exciton optical transitions. 
 The debate on the electronic structure at the experimental level, calls for accurate first principles calculations and advocates for the development of approaches that can capture subtle physical effects.  In this context it is relevant to look at the perfomance for the electronic structure of QS$GW$, which is a nonpertubative method, thus independent of the DFT starting point, and of QS$GW$+BSE for the optical properties.
\begin{figure}[h!]
\includegraphics[width=0.34\textwidth,clip=true,trim=0.0cm 0.0cm 0.0cm 0.0cm]{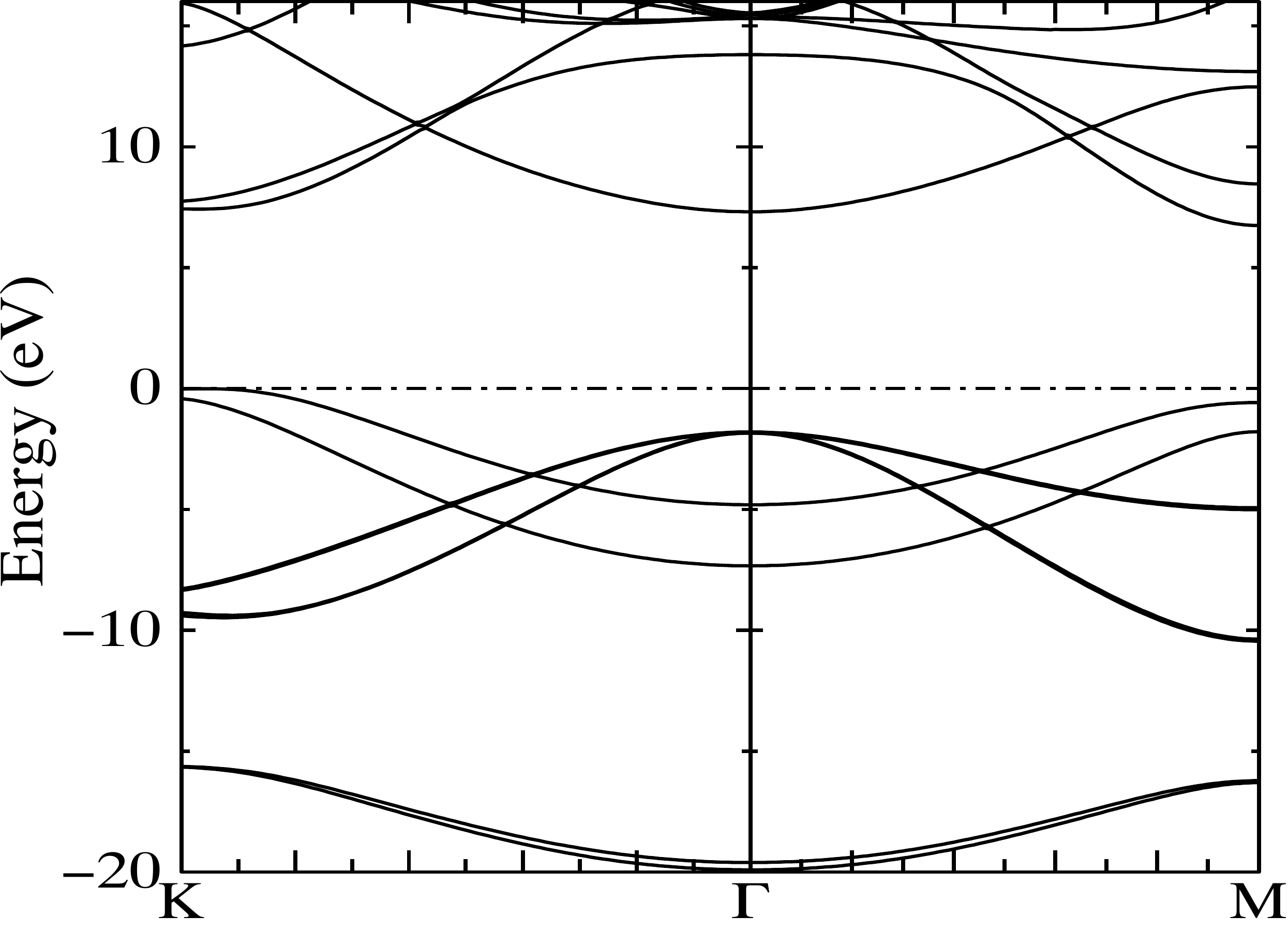}
\caption{QS$GW$ Band structure for {\it h}-BN along the  $K\,\Gamma\,M$ high-symmetry direction.}
\label{fig:hbn_bs}
\end{figure}


Figure~\ref{fig:hbn_bs} presents the QS$GW$ band structure along the $K\,\Gamma\,M$ high-symmetry direction.  The fundamental band gap is indirect and the maximum in the valence band occurs at about 95\% of the way along the line joing $\Gamma$ and K -- as in Ref.~\onlinecite{hbn_bs}.  The value for the fundamental LDA band gap (4.05~eV) is corrected by 2.18~eV at the $G_0W_0(Z =1)$ level. Self-consistency further opens the gap to 6.74~eV. The same trend is observed for the direct gap.

Results at the LDA and $G_0W_0$ level are in agreement with previous works\cite{PhysRevB.87.035404,PhysRevB.64.035104,PhysRevLett.96.026402}. Regarding the self-consistency, interestingly the value found by QS$GW$ falls in between the values for $GW_0$ (energies updated in $G$) and $GW$ (energies updated in both $G$ and $W$)  reported for instance in Ref.~\onlinecite{PhysRevB.87.035404}. Because of the spread of values mentioned above, comparison with experiment is difficult. As an example, table 1 of Ref.~\onlinecite{SOLOZHENKO20011331} summarises experimental values for the bandgap which range from 3.6 to 7.1~eV. Recent studies agree more closely with values between 6.1~eV~\cite{Cassabois2016} and 6.4~eV~\cite{hBN_BE}, consistent with the $G_0W_0$ results in this work and in the literature.\cite{PhysRevB.87.035404,PhysRevB.64.035104,PhysRevLett.96.026402} As discussed above, the QS$GW$ overestimates the bandgap accounting for the missing vertex corrections and electron-phonon interaction. The latter is predicted~\cite{PhysRevLett.101.106405} to be of the order of 0.1~eV.

\begin{figure}[h!]
  \includegraphics[width=0.38\textwidth,clip=true,trim=0.0cm 0.0cm 0.0cm 0.0cm]{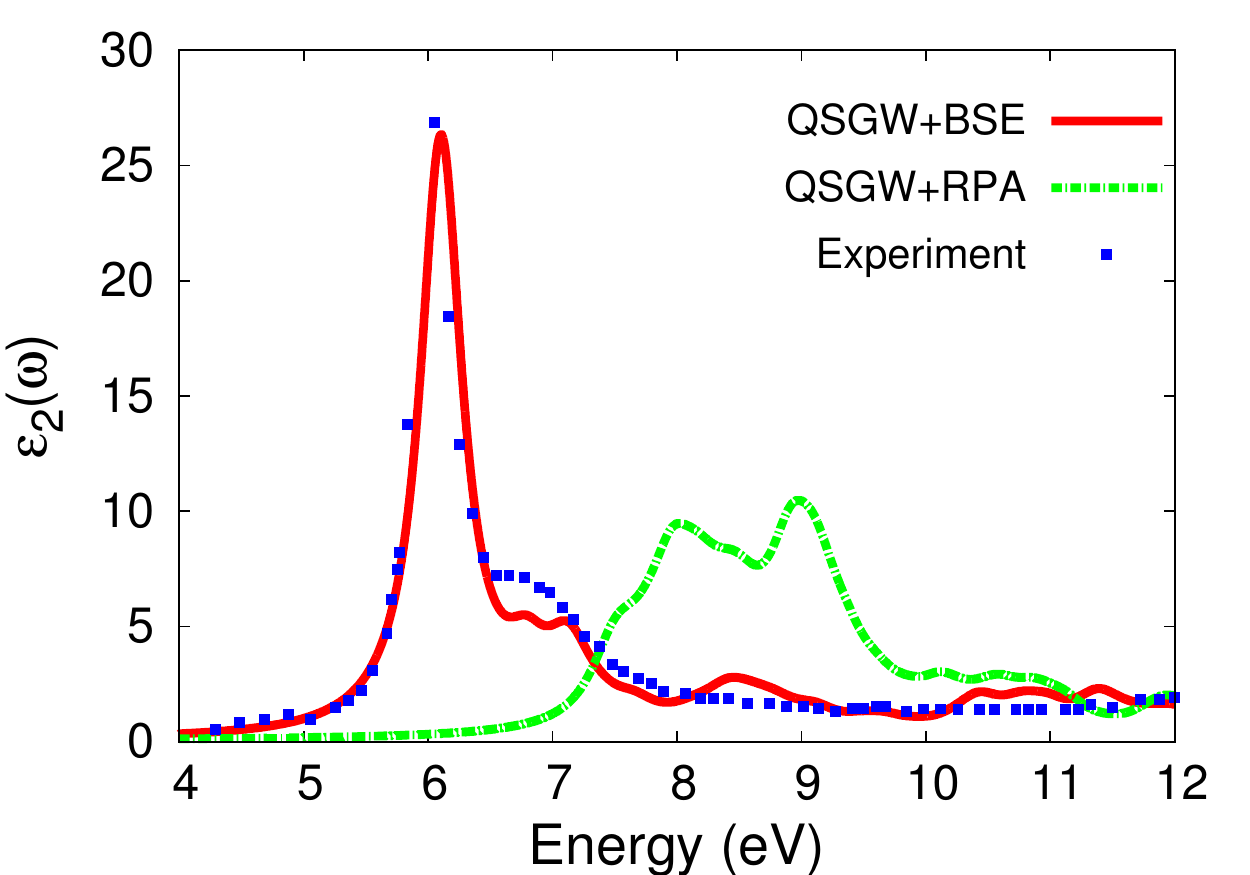}
\caption{Imaginary part of the macroscopic dielectric function for bulk hexagonal BN. The light is polarized parallel to the layers. Theoretical results at the QS$GW$+RPA (green line) and  QS$GW$+BSE (red line) level are compared with the experimental data\cite{hbn_exp}(blue squares).}
\label{image:hbn}
\end{figure}

Figure~\ref{image:hbn} compares the absorption spectrum (QS$GW$+RPA and QS$GW$+BSE) with the experimental spectrum.\cite{hbn_exp} As known from the literature (see e.g. Ref.~\onlinecite{hbn_marini}) including excitonic and local-field effects remarkably improves the agreement with experiment for this compound. The spectrum obtained shows a strong bound exciton in very good agreement with the results in the literature obtained at a similar level of theory~\cite{hbn_marini,PhysRevLett.100.189701,PhysRevLett.96.026402}. Previous works at $GW$+BSE level agree on 0.7~eV exciton binding energy, which is remarkably higher than values inferred from experiments: Refs.~\onlinecite{Watanabe2004,Cassabois2016} infer a binding energy of 130--149~meV from photoluminescence experiments; Ref.~\onlinecite{hBN_BE} obtains instead a value of 380~meV by combining photoluminescence with photoconductivity. The large discrepancy between first-principles and experiment can be partially attributed to temperature effects which are found to reduce exciton binding energy by 30\%.~\cite{PhysRevLett.101.106405} 
In this work, by comparing the  QS$GW$+RPA and QS$GW$+BSE, we obtain a value of 1.2~eV, largely overestimated with respect to other theoretical values. As discussed for LiF, the overestimation is due to missing vertex corrections which lead to an underscreened $W$. While in LiF the errors in the bandgap and binding energy cancel out only partially, for {\it h}-BN cancellation of errors gives a very good agreement with the experiment---while theoretical results at the level of $G_0W_0$+BSE are usually underestimating the exciton position by $0.2-0.3$~eV.~\cite{PhysRevLett.100.189701,PhysRevLett.96.026402,2005cond.mat..8421W} The difference of performance of QS$GW$+BSE for the two compounds may be due to the reduction of the exciton binding energy in {\it h}-BN with temperature mentioned previously.~\cite{PhysRevLett.101.106405}


\subsection{Germanium}
\label{ss:Ge}
Germanium is a semiconductor with an indirect band gap between $\Gamma$ and $L$ of about $0.7$~eV~\cite{ge_paper,Ge_gap2}. The direct $\Gamma$ bandgap is about $0.9$~eV and the valence band splitting due to spin-orbit coupling at $\Gamma$ is about 0.29~eV.\cite{GROSSO2000230} The interest on Ge for applications in devices (as for example for germanium-on-silicon lasers~\cite{Liang2010}) advocates the development of accurate and reliable approaches to study both the electronic structure and optical properties. 

Figure~\ref{image:ge_bands} presents the (spin-unpolarized) LDA and (spin-unpolarized and spin-polarized) QS$GW$ band structures for Ge. The QS$GW$ correctly predicts a fundamental gap between $\Gamma$ and $L$ of $0.78$~eV and a $1.09$~eV bandgap at $\Gamma$, so overestimating both the gaps and the energy differences between the two conduction valleys with respect to the experiment. The splitting of the QS$GW$ valence bands when including spin-orbit coupling is 0.3~eV; in agreement with the value quoted in reference~\onlinecite{GROSSO2000230}. The LDA predicts the wrong ordering of the valley in the bandstructure: at $\Gamma$ the conduction band is degenerate with the heavy and light hole bands. The split-off band, which is expected to be degenerate with heavy and light hole bands when no spin-orbit interaction is included, is split by $0.13$~eV. Furthermore, the curvature of both the conduction and split-off band is remarkably larger with respect to the QS$GW$. Note that $G_0W_0 (Z=1)$ (not shown) provides the correct ordering of the bands at $\Gamma$ and a direct bandgap of $0.96$~eV. The failure of LDA (and GGA) to predict the correct ordering of the conduction valleys has been already extensively discussed in the literature (see e.g Refs~\onlinecite{semicore1,Ge_gap2,ge_paper,SC_BE}). Within the pseudopotential approach, available pseudopotentials with $d$ electrons in the core give, by virtue of error cancellation, a semiconductor with the correct band-ordering, though the bandgap is underestimated. When semicore electrons are considered in the pseudopotential (or core corrections considered), the all electron picture is usually recovered. The effect of the pseudopotential, and specifically the effect of semicore states, has been studied in previous works also in connection with the $GW$ approximation and self-consistency.\cite{semicore1,Ge_gap2,ge_paper,ge_comm}

{\begin{figure}[h!]
\includegraphics[width=0.30\textwidth,clip=true,trim=0.0cm 0.0cm 0.0cm 0.0cm]{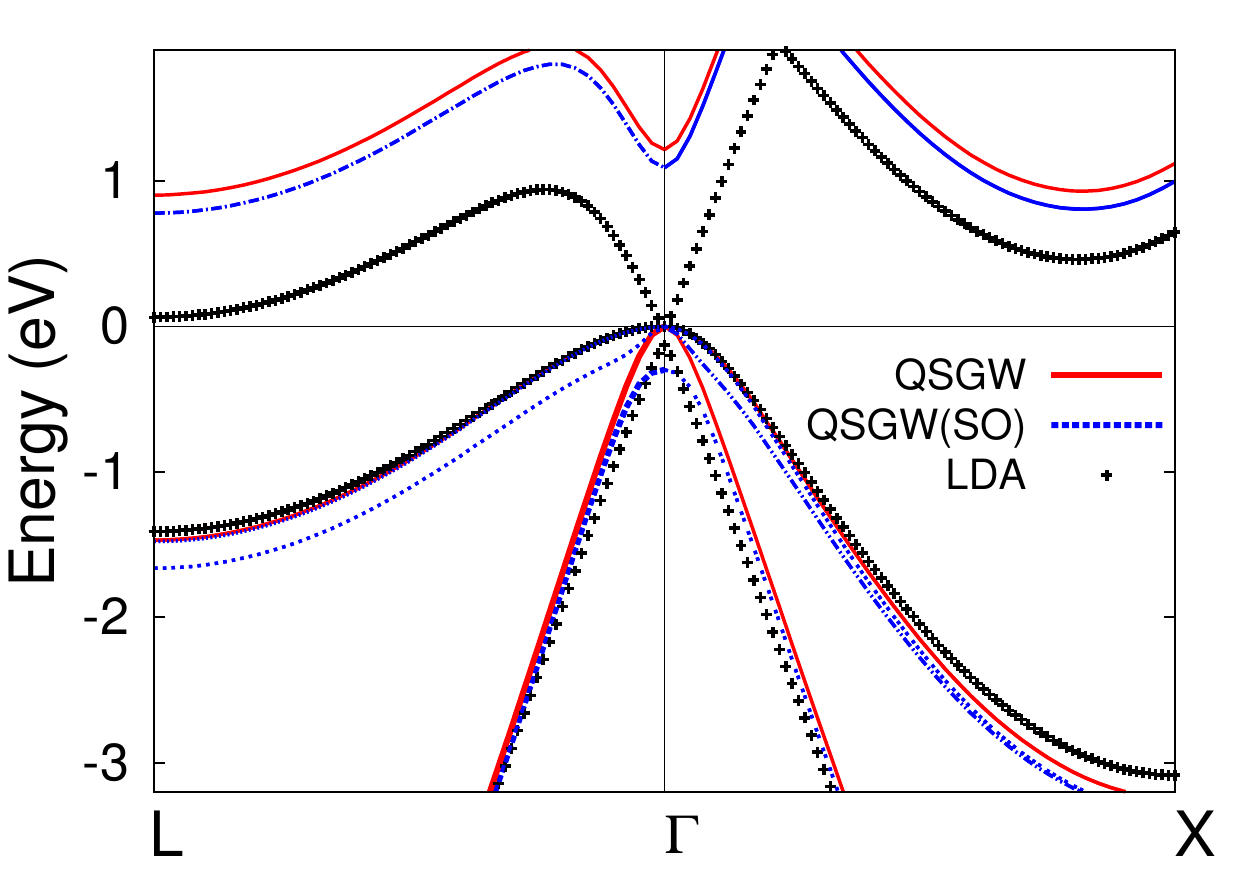}
\caption{LDA (dots) and QSGW, with (dashed line) and without (continuous line) spin-orbit coupling, band structures of bulk Germanium along the L-$\Gamma$-X directions in the Brillouin zone.}
\label{image:ge_bands}
\end{figure}}


When calculating the dielectric function from methods relying on perturbative corrections of the LDA and GGA electronic structure such as $G_0W_0$ there two main issues stemming from the zero-gap prediction of LDA/GGA: the overscreening of the $W$ (already observed  e.g. in LiF and that partially cancels with other missing effects), and the calculation of the TDMEs when using the usual rescaling by DFT energies as in Sec.~\ref{subsec:ome}, which in this case are zero/negative. Here, the first issue is addressed by using the QS$GW$, the second by calculating the contribution from the nonlocal potential to the TDMEs explicitely as in Eq.~\ref{eq:dsigdp}. Figure~\ref{image:ge} then presents the real and imaginary parts of the macroscopic dielectric function for Ge, with the TDMEs calculated using Eq.~\ref{eq:dsigdp}. Both position and intensities of the main features are well reproduced when comparing with experiment.\cite{ge_exp}

\par
{\begin{figure}[h!]
\includegraphics[width=0.38\textwidth,clip=true,trim=0.0cm 0.0cm 0.0cm 0.0cm]{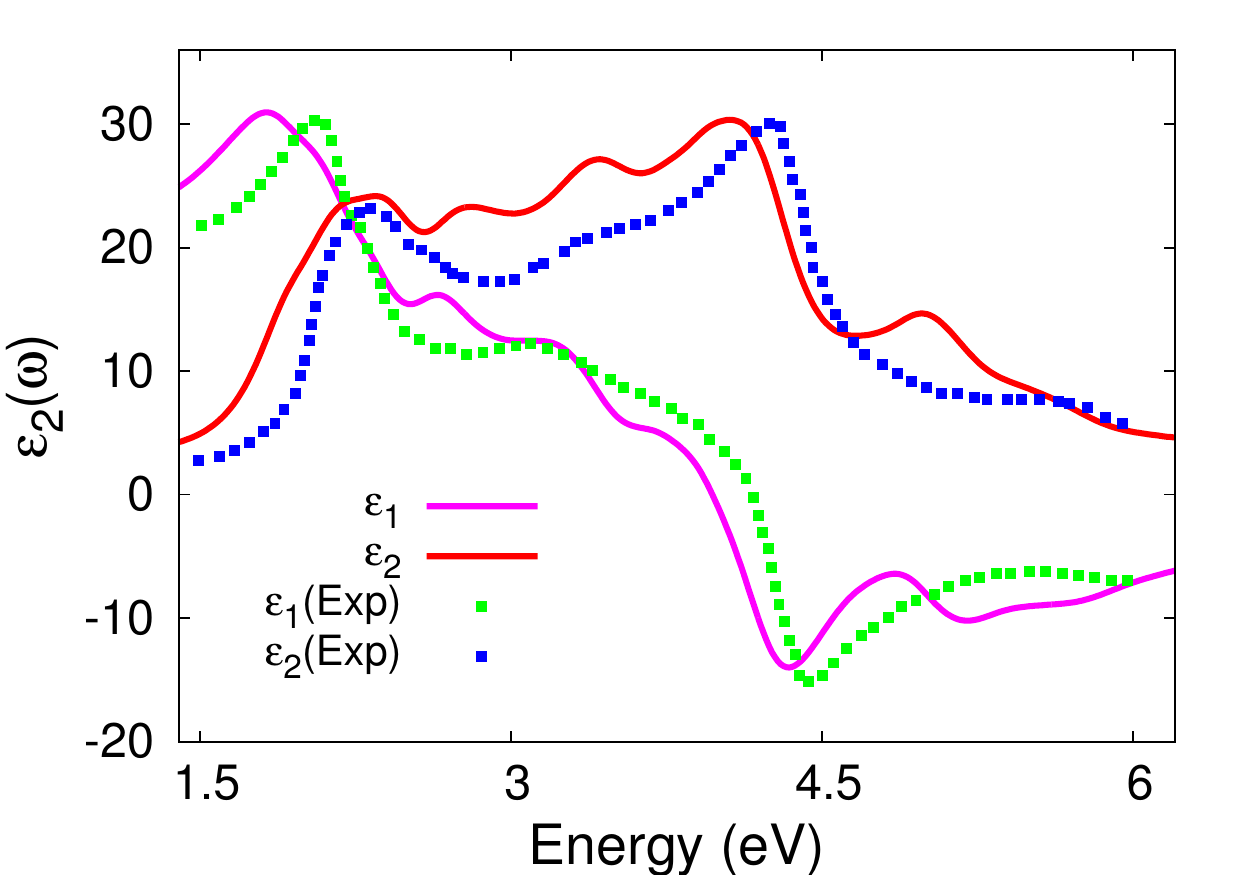}
\caption{Real and imaginary parts of the macroscopic dielectric function for bulk Ge calculated using QS$GW$+BSE (continuous line) is compared with the experimental results (squares).}
\label{image:ge}
\end{figure}}

\subsection{Nickel Oxide}
\label{ss:nio}
The transition metal oxide NiO is an antiferromagnetic material with a magnetic moment of $1.9~\mu_{\rm B}$ and a band gap of 4.3~eV.\cite{QSGW_prl1,NiO_exp} It is a prototypical strongly correlated material, i.e. a material for which one-particle approaches fail to describe even qualitative features. In particular, NiO belongs to $3d$ transition-metal compounds for which DFT predicts a metal/semiconductor rather than a wide-gap insulator. The error has been traced down to the inability of one-particle approaches to capture the correlation effects of $d$ electrons.~\cite{PhysRevLett.55.418}\par
Consistently with this picture and results reported previously,~\cite{QSGW_paper,QSGW_PRL} the LDA band gap is found to be about 0.4~eV. The $G_0W_0$ calculated band gap is opened to 1.7~eV ($Z=1$) which is, as expected, still heavily underestimated. In fact, the LDA and $GW$ (with eigenvalue-only self-consistency) bandstructures have been thoroughly analyzed in Ref.~\onlinecite{QSGW_paper}. It was found that for both approaches, the conduction band dispersions are qualitatively wrong and the conduction band minimum is not at $\Gamma$, when compared with QS$GW$.    
Selfconsistency at the QS$GW$ level gives  an indirect bandgap from U to $\Gamma$ of 4.86~eV (direct gap of 5.56~eV at $\Gamma$), overestimated by about 0.5~eV. As previously discussed the main sources of the difference with the experimental value can be traced back to vertex and temperature effects.  
Beside the bandgap, the magnetic moment is also severely underestimated in the LDA; $1.23~\mu_{\rm B}$ versus $1.71~\mu_{\rm B}$ at QS$GW$.

Figure~\ref{image:nio} presents the calculated absorption spectrum at the QS$GW$+BSE level. Because of the large errors in the calculated electronic structure, any perturbative approach starting from the LDA, such as $G_0W_0$+BSE, is expected to poorly predict the optical absorption spectrum.    
Regarding the treatment of the TDMEs, since LDA gives qualitatively wrong results and the  QS$GW$ eigenfunctions differ significantly from the LDA, using the LDA energies as the scaling factor leads to poor results. Alternative schemes, such as the $\partial\Sigma/\partial p$ scheme, are in this case mandatory. Indeed, the all-electron QS$GW$+BSE with $\partial\Sigma/\partial p$ produces a spectrum in very good agreement with experiment for NiO,~\cite{NiO_expopt}, but for a shift of about $1$~eV in the spectral onset due to the overestimation of the bandgap, only partially compensated by the error in the binding energy. The agreement with the experiment is visibly better than at the $GW$+RPA level where the onset is overestimated by over $2$~eV and the intensity overestimated.

{\begin{figure}[h!]
\includegraphics[width=0.38\textwidth,clip=true,trim=0.0cm 0.0cm 0.0cm 0.0cm]{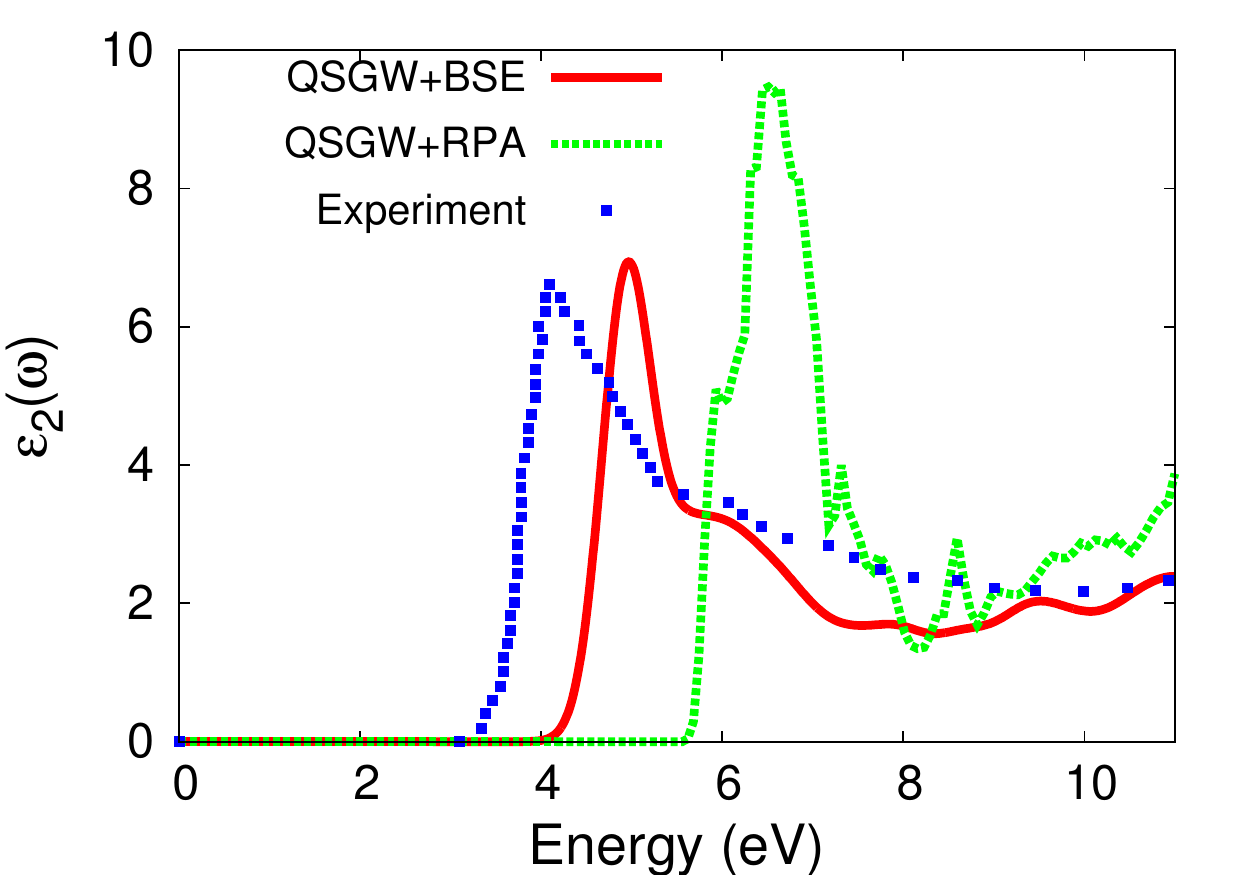}
\caption{Imaginary part of the macroscopic dielectric function for NiO.  The experimental data (blue dots)\cite{NiO_expopt} is presented along with spectra calulated at the level of QS$GW$+BSE (red line) and QS$GW$+RPA (green line).}
\label{image:nio}
\end{figure}}

%
%
\section{Conclusions}
We have combined the QS$GW$ approach for calculating the electronic structure with the solution of the BSE for calculating the optical spectrum. The macroscopic dielectric function of LiF, Si, {\it h}-BN, Ge and NiO have been computed with this approach (QS$GW$+BSE) and compared with the $G_0W_0$+BSE---which is commonly used for calculating optical absorption of materials---and with the QS$GW$+RPA. The comparison with the latter approach highlights the need of including excitonic effects, as already extensively discussed in the literature (see e.g. Ref.~\onlinecite{onida_electronic_2002}). The comparison of QS$GW$+BSE and $G_0W_0$+BSE instead highlights the merits and limits of the QS$GW$ for calculating the electronic structure.

For Si, LiF and {\it h}-BN, the performance of the two methods is similar. More specifically a slight improvement is found for Si and for {\it h}-BN, while in LiF the exciton position is blue-shifted by almost 1~eV. These results have been rationalised by considering the error cancellation which is usually responsible for the good agreement of the $G_0W_0$ calculated bandgap with the experimental gap. Namely, the $W$ is calculated within the RPA (overestimation), using as input DFT energies (underestimation). The QS$GW$ corrects for the underestimation from the DFT energies, but $W$ is still calculated within the RPA. Furthermore as it emerged from recent literature, the neglection of electron-phonon interaction leads to a bandgap overestimation of the order of hundreds of meV.         

The benefits of the present approach have been made clear for Ge and NiO. For different reasons, LDA is not a good starting point for both those systems. For Ge, a narrow-gap semiconductor, the bandgap is inverted. The $G_0W_0$ partially corrects the bandgap. The more severe problem is though that the bandgap rescaling, which accounts for the nonlocal contribution to the transition dipoles, cannot be applied. For NiO, a strongly correlated transition metal oxide, the LDA+$G_0W_0$ severely underestimates the fundamental bandgap, and a better starting point, such as that provided by QS$GW$ is essential to get the electronic structure and as a consequence the optical properties. 

To summarise, the key advantage of the approach here presented over the more standard $GW$+BSE is the possibility of calculating the optical properties of materials for which $GW$ on top of the standard DFT provides a poor description of the electronic structure. Furthermore, as we employ an all-electron basis, we eliminate the dependence on the choice of the pseudopotential which sometimes---though it should not be the case---can substantially influence the $GW$ results.\footnote{It should be noted that most of the pseudopotentials are designed for ground-state, rather than excited state properties}

The overestimation of the bandgap, and thus of the spectrum onset, observed for wide-band gap insulators such as LiF, and for NiO, draws the attention on important effects missing from the present framework. In particular, the RPA for $W$ is clearly insufficient when the accurate electronic structure is used rather than the DFT one, and one would need to introduce a BSE-like vertex correction to $W$.~\cite{PhysRevLett.99.246403} Further, the electron-phonon and exciton-photon interactions also play an important role and would need to be included when aiming at accurate predictions of materials optical properties.  


\begin{acknowledgments}
  The authors would like to thank all those involved in the {\it CCP flagship project: Quasiparticle Self-Consistent $GW$ for Next-Generation Electronic Structure}, especially Scott Mckechnie for his help. MG acknowledges Maurizia Palummo, Daniele Varsano and Claudio Attaccalite for discussion on the {\it h}-BN bandgap. We are grateful for support from the Engineering and Physical Sciences Research Council, under grant EP/M011631/1. MvS was supported in part by the Simons Foundation.
\end{acknowledgments}
\bibliography{references}
\end{document}